\documentclass[aps,pre,reprint,longbibliography,amsmath,amssymb,floatfix,raggedbottom]{revtex4-2}
\usepackage{bm,graphicx}
\IfFileExists{stmaryrd.sty}{\usepackage{stmaryrd}}{%
  \newcommand{\llbracket}{\mathopen{[\![}}%
  \newcommand{\rrbracket}{\mathclose{]\!]}}%
}
\usepackage[colorlinks=true,allcolors=blue]{hyperref}
\newcommand{\dd}{\mathrm d}
\newcommand{\epsc}{\epsilon_{21}}
\newcommand{\sigc}{\sigma_{21}}
\newcommand{\Ma}{\mathrm{Ma}}
\newcommand{\Mac}{\mathrm{Ma}_c}
\newcommand{\ez}{\boldsymbol e_z}
\newcommand{\Pp}{\boldsymbol P}
\newcommand{\Pperp}{P_\perp}
\newcommand{\Om}{\boldsymbol\Omega}
\newcommand{\I}{\boldsymbol I}
\newcommand{\PiR}{\boldsymbol\Pi}
\newcommand{\order}{\mathcal O}
\begin{document}
\title{Electrohydrodynamic interactions of two Quincke rotors: asymptotic theory and boundary element simulations}
\author{Debasish Das}
\affiliation{Department of Mathematics and Statistics, University of Strathclyde, Glasgow, United Kingdom}
\begin{abstract}
Weakly conducting spheres suspended in a leaky dielectric fluid can
undergo spontaneous Quincke rotation when the applied electric field
exceeds a critical value. We investigate how electric and hydrodynamic
interactions modify this instability and the subsequent rotation of two
spheres at fixed separation and orientation. Starting from electric
dipole and hydrodynamic rotlet interactions, we develop a systematic
far-field asymptotic theory. Away from onset, slow drift of the rotation
axes makes a regular expansion fail at long times. A multiple-scale
expansion gives slow phase equations that select the axes according to
pair orientation and field strength. Near onset, weak interactions
produce corrections comparable to the isolated rotation magnitude.
A detuned multiple-scale expansion gives coupled equations for the
transverse dipole amplitudes. Their linearization identifies four
collective rotation modes, whose critical fields and steady branches
with one active mode are determined together with their stability.
At leading order near onset, electric interaction shifts all four modal
thresholds equally, whereas hydrodynamic interaction makes
counter-rotation appear first for pair angles below
$\cos^{-1}(1/\sqrt3)\simeq54.7^\circ$ and co-rotation above it.
Away from onset, the angle separating stable counter-rotation from
stable co-rotation increases with field strength. Boundary element method (BEM)
simulations place onset within a field interval for each configuration
and resolve the growth and relaxation of rotation. Steady-speed
comparisons show how pair orientation changes the balance between
electric and hydrodynamic effects. We compare the unexpanded dipole-rotlet model of Das and Saintillan [Phys. Rev. E 87, 043014 (2013)] with our asymptotic theory and BEM simulations.
\end{abstract}
\maketitle
\raggedbottom

\section{Introduction}
Electric fields provide a means of controlling the organization, transport,
and mechanical response of particles suspended in fluids. Particle motion is governed by electric forces and torques and by their
coupling to the surrounding fluid. In electrorheological suspensions,
anisotropic polarization interactions promote chains and columns whose
resistance to deformation produces a field-dependent yield
stress~\cite{Winslow1949,Halsey1992,ParthasarathyKlingenberg1996}.
Their response under flow depends on the competition between electric
and hydrodynamic forces and the resulting particle
rearrangements~\cite{Klingenberg1989,KlingenbergZukoski1990,Klingenberg1991,BonnecazeBradyDynamic1992,BonnecazeBradyYield1992};
structural relaxation also contributes to their oscillatory mechanical
response~\cite{Klingenberg1993}. Nonuniform fields instead exert a net
dielectrophoretic force~\cite{Pohl1951}, which has been used to characterize,
separate, position, and pattern biological cells and to manipulate smaller
biological objects~\cite{Gagnon2011,Pethig2017}. These applications exploit
the dependence of polarization on material properties, field frequency,
and the electrical response of the suspending medium.

Electric fields can also drive flow in the surrounding fluid. In induced-charge
electro-osmosis, an applied field acts on the diffuse charge that it induces
near a polarizable surface, producing fluid motion whose characteristic
speed is quadratic in the field at weak forcing~\cite{BazantSquires2004,SquiresBazant2004}.
Breaking the symmetry of this flow can produce fluid pumping or propel a
suspended particle~\cite{SquiresBazant2006}, as demonstrated by the electrophoresis
of metallodielectric Janus particles~\cite{GangwalICEP2008}. Surface
heterogeneity also permits field-controlled assembly and changes in
collective organization~\cite{GangwalAssembly2008,Yan2016,Zhang2017}.
Particle shape and nearby electrodes further modify propulsion,
hydrodynamic trapping, and electric orientation or
rotation~\cite{Boymelgreen2022,Katuri2022,Sofer2023}. Although the charge
distributions and forcing protocols differ among these systems, they share
a coupling between polarization, interfacial charge transport, and the
motion of the surrounding fluid.

Here we focus on leaky dielectric particles, for which finite electrical
conduction and charge relaxation are essential. The Taylor--Melcher
description treats the bulk phases as electrically neutral, with
conduction obeying Ohm's law, while allowing free charge to accumulate at their
interface~\cite{MelcherTaylor1969,Saville1997}. The interfacial charge is
governed by the balance of conduction currents and surface convection;
the associated Maxwell stress couples the electric field to the fluid
motion. For deformable drops, tangential electric stresses drive
interfacial circulation, which can redistribute surface charge and modify
the drop deformation. For rigid particles, electric forces and torques can
drive translation and rotation, coupling particle motion to charge
relaxation. Rotation redistributes surface charge, while translation
through a nonuniform electric field changes the field experienced by the
particle. This feedback becomes important when charge relaxation and
particle motion occur on comparable time scales~\cite{Saville1997,DasSaintillan2013}.

Quincke rotation is a particularly direct manifestation of this feedback:
an initially stationary, isotropic particle spontaneously rotates in a
uniform, time-independent electric field~\cite{Quincke1896,Turcu1987,Jones1984}. When charge
relaxes more slowly inside the particle than in the fluid, the relaxing
part of the induced dipole is antiparallel to the imposed field. A small
perturbation tilts the relaxing component of the induced dipole, generating
an electric torque that can amplify the perturbation. Above a critical field, the feedback between
rotation and surface-charge convection destabilizes the nonrotating
state. The resulting steady rotation is sustained by electric torque
balancing viscous resistance. In the ideal inertialess problem
for an isolated sphere, the rotation axis lies perpendicular to the field
but its azimuth is arbitrary. With rotational inertia, the dynamics can
be richer: experiments on insulating rotors have demonstrated transitions
from steady rotation to chaotic motion, including Lorenz-type
chaos~\cite{Lemaire2002,Peters2005}.

Quincke rotation also affects suspension rheology and electrical
transport. Experiments show that it can reduce the apparent viscosity
by assisting an imposed shear~\cite{Lobry1999}, while weak shear can
reorient the rotation axes and alter the low-shear
response~\cite{Lemaire2008}. The competition between shear, electric
attraction, and rotationally generated flow can produce layering and
phase separation~\cite{PannacciRheology2007}. Viscosity reduction also
increases capillary flow rates and changes pressure-driven velocity
profiles~\cite{Lemaire2006,Peters2010}. Although theory predicts negative
apparent viscosity in suitable regimes~\cite{Cebers2004}, the
pressure-driven experiments found a reduction without reaching negative
viscosity~\cite{Peters2010}. Continuum theories coupling particle
rotation to the fluid predict polar ordering and spontaneous flows,
including flow below the isolated-particle critical
field~\cite{Belovs2014,Belovs2020}. Quincke rotation also
enhances the electrical conductivity of the suspension~\cite{PannacciConductivity2007,Pannacci2009}.
These effects connect the rotational instability to suspension
microstructure and macroscopic transport in the bulk.

Near a solid boundary, hydrodynamic coupling converts Quincke rotation
into translation, producing self-propelled colloidal rollers. These
particles continually convert externally supplied electrical energy into
motion and have become a model system for investigating active matter.
Their motion and interactions can be described through the electric
potential and fluid flow, governed by Laplace's and Stokes equations and
coupled through surface-charge dynamics. Populations of spherical rollers
form traveling bands and macroscopically directed polar fluids through
interactions mediated by the electric field and the surrounding
fluid~\cite{Bricard2013}. Confinement organizes this motion into persistent
vortices~\cite{Bricard2015}. Disorder can redirect a flock into sparse
channels and eventually destroy polar order~\cite{Morin2017}, whereas
obstacles in an annular track can induce repeated flow reversals or arrays
of counter-rotating vortices~\cite{ZhangObstacles2020}.

The electrical driving also provides control over the persistence and
collective dynamics of the motion. Pulsing the field allows charge to
relax between runs, producing tunable run-and-tumble trajectories and
dynamic clustering~\cite{Karani2019}. In larger populations, pulsed
driving supports spontaneous shock waves whose fronts propagate faster
than the particles they entrain~\cite{ZhangShock2023}. Attraction generated
by electrohydrodynamic flow can compete with Quincke propulsion: in
experiments on small spherical rollers, initially compact crystallites
become motile and continually reshape, merge, and split before dispersing
into an active gas as the field increases~\cite{Mauleon2020}. The field
therefore controls several coupled processes, including propulsion,
charge memory, and the interactions responsible for aggregation.

Particle shape introduces further ways of converting rotation into
transport. Calculations for asymmetric rigid particles show that Quincke
rotation can generate propulsion in a bulk fluid, without a nearby
wall~\cite{DasLauga2019}. Experiments with pear-shaped rollers reveal
field-dependent curved trajectories and collective states ranging from
vortices to rotating flocks~\cite{ZhangChiral2020}; their chiral motion can
also suppress long-wavelength density fluctuations while preserving
structure at intermediate scales~\cite{ZhangHyperuniform2022}. Quincke
dumbbells exhibit spinning and orbital motion, and collisions can bind
them into rotating tetramers and larger clusters~\cite{MauleonAmieva2023}.
Magnetic control offers an additional independent handle: magnetic
Quincke rollers form active chains and can be steered or confined by
magnetic fields~\cite{ReyesGarza2023}, while other magnetic particles
exhibit coexisting rolling directions parallel and perpendicular to the
field~\cite{Fitzgerald2025}.

Recent experiments also extend the dynamics beyond steady spinning.
Small colloids near electrodes can undergo sustained Quincke oscillations,
for which ion transport and the spatial variation of conductivity near
the electrode are essential~\cite{ZhangOscillations2021}. Interactions
between these oscillators produce simultaneous phase synchronization and
alignment of the oscillation axes, with different collective organization
in dilute suspensions and dense clusters~\cite{Leyva2026}. In a distinct
realization with spatially anchored, three-dimensionally printed Quincke
micromotors, interactions organize neighboring motors into alternating
rotation directions, coherent precession, and propagating phase
waves~\cite{Braun2026}. These systems highlight the importance of coupled
rotational dynamics, although their oscillation or precession phases are
different from the azimuthal orientation of the steady rotating state of
an isotropic sphere.

A common ingredient in these diverse collective behaviors is the
interaction between electrically driven particles, mediated by both the
electric field and the surrounding fluid. Each particle perturbs the
local electric field and fluid vorticity experienced by its neighbors,
modifying their rotation speeds, selecting their rotation axes, and
shifting the critical field for instability. Near a solid boundary,
additional electric and hydrodynamic interactions with the surface
complicate this coupling. An unbounded fluid therefore provides a simpler
starting point for establishing the particle--particle interaction physics.

Previous work developed a far-field model for two Quincke spheres that
captured geometry-dependent changes in onset and coupled rotational and
translational dynamics~\cite{DasSaintillan2013}. Its self-consistent coupled
equations provide useful predictions and retain selected higher-order
interaction contributions, but these do not constitute a complete
asymptotic expansion at those higher orders. We refer to these coupled
equations as the unexpanded dipole--rotlet model.
Direct numerical simulations have also resolved hydrodynamic interactions
between Quincke rollers near a substrate, including lubrication
effects~\cite{Imamura2023}. Those calculations represent electrical
interactions through dipoles and prescribe the Quincke rotation speed
from an isolated-particle steady solution.

Here, we study two identical spheres at a fixed separation and orientation
in an unbounded fluid subject to a uniform dc electric field. Freezing
the pair geometry isolates the local rotational coupling relevant to a
suspension. Starting from electric dipole and hydrodynamic rotlet
interactions, we develop the first systematic multiple-scale asymptotic
theory of the coupled rotational dynamics of two particles undergoing
Quincke rotation, covering regimes both near and away from onset.
Two effects limit a regular far-field expansion. Because every rotation
axis perpendicular to the field is equally possible for an isolated
sphere, weak interactions can turn the axes slowly. A correction
calculated with fixed axes then grows in time. Independently, the decay
rate of rotation-magnitude perturbations vanishes at onset, so the
interaction correction can cease to be small even for widely separated
particles.

Away from onset, a multiple-scale expansion with a fast and a slow time
gives phase equations describing the evolution and selection of the
rotation axes. Near onset, we take the distance from the isolated
critical field to be of the same order as the interaction strength
$R^{-3}$ and expand about the critical non-rotating state.
The resulting detuned multiple-scale expansion gives coupled amplitude
equations that determine the critical fields, nonlinear saturation,
and stability of competing rotation patterns. We separate the electric
and hydrodynamic contributions to these dynamics.

We also develop a new boundary-element formulation for two interacting
spheres undergoing Quincke rotation. This extends the boundary element
method (BEM) previously used for isolated drops~\cite{DasSaintillan2017}
and isolated rigid particles~\cite{DasLauga2019} to the coupled
two-particle problem. The simulations evolve surface charge on both
spheres and resolve their electric and hydrodynamic interactions through
the coupled boundary-value problems. They provide a numerical benchmark
for the asymptotic theory and the unexpanded dipole--rotlet model at
finite separation.

Section~\ref{sec:setup} describes the problem set-up, governing
equations and dimensionless parameters. Sections~\ref{sec:regular}--\ref{sec:breakdown} develop the regular
expansion and slow phase equations away from onset, then establish
their near-onset limitation. Sections~\ref{sec:multiple}--\ref{sec:switched}
derive the near-onset amplitude equations and their rotation patterns,
critical fields, and interaction mechanisms. Section~\ref{sec:finiteamp}
returns to steady speeds away from onset, before
Secs.~\ref{sec:bem} and \ref{sec:numerics} present the BEM formulation
and comparisons.

\section{Problem set-up}
\label{sec:setup}
\begin{figure}[!t]
\centering
\includegraphics[width=\columnwidth]{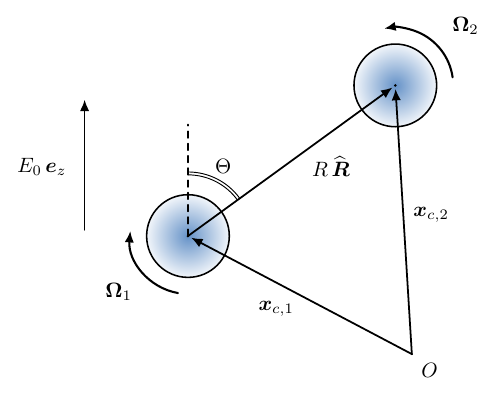}
\caption{Geometry of two identical Quincke rotors in a uniform dc electric
field $E_0\ez$. The centres are located at $\boldsymbol x_{c,1}$ and
$\boldsymbol x_{c,2}$, with separation vector $R\widehat{\boldsymbol R}$
and lengths scaled by the sphere radius $a$. The pair angle $\Theta$ is
the angle between the line of centres and the external electric field. The curved arrows illustrate
a counter-rotating state.}
\label{fig:setup}
\end{figure}

We consider two identical rigid spheres of radius $a$
in an unbounded weakly conducting Newtonian fluid subject to a uniform dc electric field $E_0\ez$.
The geometry is illustrated in Fig.~\ref{fig:setup}. To isolate rotational coupling, we freeze the pair separation and
orientation while the surface charge and angular velocities evolve.
In a suspension, particle translation would also change the pair
configuration~\cite{das2023absence}. The hydrodynamic calculation is
performed with the electric torques prescribed and the net forces on
both spheres set to zero. The resulting translational velocities are discarded. The fluid has
permittivity $\epsilon_1$, conductivity $\sigma_1$, and viscosity $\mu$;
the particle properties are $\epsilon_2$ and $\sigma_2$. We neglect
inertia, bulk free charge, and surface conduction.

The material contrasts and Maxwell--Wagner relaxation time are
\begin{equation}
\epsc=\frac{\epsilon_2-\epsilon_1}{\epsilon_2+2\epsilon_1},\;
\sigc=\frac{\sigma_2-\sigma_1}{\sigma_2+2\sigma_1},\;
\tau_{\mathrm{MW}}=\frac{\epsilon_2+2\epsilon_1}{\sigma_2+2\sigma_1}.
\label{eq:contrasts_tauMW}
\end{equation}
Lengths, time, electric fields, dipole coefficients, and angular velocities
are scaled by $a$, $\tau_{\mathrm{MW}}$, $E_0$, $a^3E_0$, and
$\tau_{\mathrm{MW}}^{-1}$, respectively. For sphere $i$, we denote the
dimensionless dipole coefficient, local electric field, and angular
velocity by $\Pp_i$, $\boldsymbol E_i$, and $\Om_i$. The normalization
of $\Pp_i$ is such that the physical dipole moment is $
\boldsymbol P_i^{\mathrm{phys}}=4\pi\epsilon_1a^3E_0\Pp_i.$
The dimensionless Mason number is
\begin{equation}
\Ma=\frac{2\mu}{\tau_{\mathrm{MW}}\epsilon_1E_0^2}.
\label{eq:Ma}
\end{equation}
The Mason number compares the viscous stress $\mu/\tau_{\mathrm{MW}}$
with the electric stress $\epsilon_1E_0^2$.
Define the isolated critical Mason number by
\begin{equation}
\Mac=\epsc-\sigc>0.
\label{eq:Mac}
\end{equation}
The positivity condition permits Quincke rotation and is
equivalent to $\tau_2>\tau_1$, where $\tau_1=\epsilon_1/\sigma_1$ and
$\tau_2=\epsilon_2/\sigma_2$ are the bulk charge-relaxation times of the
fluid and particle. For an isolated sphere, the nonrotating state becomes
unstable when $\Ma<\Mac$, or equivalently $E_0>E_c^{(0)}$, where
\begin{equation}
E_c^{(0)}=\left[\frac{2\mu}
 {\epsilon_1\tau_{\mathrm{MW}}\Mac}\right]^{1/2}.
\label{eq:isolatedthreshold}
\end{equation}
This is the critical electric field for Quincke rotation of an isolated sphere~\cite{DasSaintillan2013}.
Interactions modify the onset field according to the pair separation
and orientation, as determined below.

Let $\boldsymbol x_{c,i}$ denote the dimensionless centre of sphere $i$.
The separation and pair-axis direction are defined by
\begin{equation}
\boldsymbol x_{c,2}-\boldsymbol x_{c,1}=R\widehat{\boldsymbol R},
\qquad |\widehat{\boldsymbol R}|=1.
\end{equation}
We consider the
far-field regime $R\gg1$ and define
\begin{equation}
\delta=R^{-3},\quad
\PiR=\I-3\widehat{\boldsymbol R}\widehat{\boldsymbol R},\quad
\cos\Theta=\widehat{\boldsymbol R}\cdot\ez.
\label{eq:geometry}
\end{equation}
Here $\Theta$ is the angle between the line of centres and the imposed
field. For identical spheres, it suffices to consider $0\leq\Theta\leq\pi/2$.

Retaining the electric dipole field and hydrodynamic rotlet disturbance gives
\begin{subequations}\label{eq:model}
\begin{align}
\boldsymbol E_i&=\ez-\delta\PiR\cdot\Pp_j,\label{eq:modelE}\\
\Om_i+\frac\delta2\PiR\cdot\Om_j
 &=\Ma^{-1}\Pp_i\times\boldsymbol E_i,\label{eq:modelO}\\
\dot{\Pp}_i+\epsc\delta\PiR\cdot\dot{\Pp}_j
 &=\Om_i\times(\Pp_i-\epsc\boldsymbol E_i)
 -(\Pp_i-\sigc\boldsymbol E_i),\label{eq:modelP}
\end{align}
\end{subequations}
where $i=1,2$, $j=3-i$, and overdots denote derivatives with respect to
dimensionless time. The term $-\delta\PiR\cdot\Pp_j$ is the electric field
produced by the dipole moment of sphere $j$, evaluated at the centre
of sphere $i$. The rotlet is the leading flow produced
by a torque; it contributes the local fluid rotation
$-(\delta/2)\PiR\cdot\Om_j$ in Eq.~\eqref{eq:modelO}.
Each sphere rotates relative to that fluid rotation at the rate set
by its electric torque. The tensor $\PiR$ has eigenvalue $-2$ along
the line of centres and $1$ in directions perpendicular to it.
The surface-charge balance and disturbance fields
are developed in Secs.~\hyperref[sec:chargeSM]{SI\,A}--\hyperref[sec:interactionSM]{C}
of the Supplemental Material (SM)~\cite{SM}; the dipole--rotlet
description follows the far-field framework of Ref.~\cite{DasSaintillan2013}.

We call Eqs.~\eqref{eq:model} the \emph{unexpanded dipole--rotlet model}.
Solving the coupled equations without expanding in $\delta$ retains
repeated mutual couplings and thereby selected contributions at higher
powers of $\delta$.

Differentiating Eq.~\eqref{eq:modelE} gives
$\dot{\boldsymbol E}_i=-\delta\PiR\cdot\dot{\Pp}_j$.
The dipole equation contains $\epsc\dot{\boldsymbol E}_i$ from the
time derivative of surface charge, as derived in
Sec.~\hyperref[sec:dimdipoleSM]{SI\,B.4} of the SM. Substitution
of the field derivative and moving this term to the left gives
Eq.~\eqref{eq:modelP}. The imposed field is constant, but the local
field changes as the other dipole evolves. Solving the two linear systems in Eqs.~\eqref{eq:modelO} and
\eqref{eq:modelP} determines the angular velocities and dipole derivatives,
giving six coupled scalar ordinary differential equations for the three
dipole components of each sphere. The matrices are invertible for
non-overlapping spheres. The finite-separation accuracy of the
dipole--rotlet approximation is assessed against BEM simulations in
Sec.~\ref{sec:numerics}.

\section{Regular expansion in the interaction strength}
\label{sec:regular}
At fixed $\Ma$ and pair orientation, write
\begin{equation}
\Pp_i=\Pp_i^{(0)}+\delta\Pp_i^{(1)}+\order(\delta^2),
\label{eq:regularP}
\end{equation}
with analogous expansions for $\boldsymbol E_i$ and $\Om_i$.
Whether the expansion remains ordered over long times
and near the isolated critical field is examined in
Secs.~\ref{sec:slowrotation} and \ref{sec:breakdown}. We substitute
these expansions into Eqs.~\eqref{eq:model} and equate coefficients of
equal powers of $\delta$. The field and torque equations first give
\begin{equation}
\boldsymbol E_i^{(0)}=\ez,\qquad
\boldsymbol E_i^{(1)}=-\PiR\cdot\Pp_j^{(0)}.
\label{eq:Eorders}
\end{equation}
\begin{equation}
\Om_i^{(0)}=\Ma^{-1}\Pp_i^{(0)}\times\ez.
\label{eq:Ozero}
\end{equation}
\begin{equation}
\begin{split}
\Om_i^{(1)}={}&\Ma^{-1}\bigl[\Pp_i^{(1)}\times\ez
 -\Pp_i^{(0)}\times(\PiR\cdot\Pp_j^{(0)})\\
&\hspace{12mm}-\tfrac12\PiR\cdot(\Pp_j^{(0)}\times\ez)\bigr].
\end{split}
\label{eq:Ofirst}
\end{equation}
The leading dipoles obey the isolated-sphere equations
\begin{equation}
\dot{\Pp}_i^{(0)}=\Om_i^{(0)}\times(\Pp_i^{(0)}-\epsc\ez)
 -(\Pp_i^{(0)}-\sigc\ez).
\label{eq:Pzero}
\end{equation}
At order $\delta$, the dipole equation gives the following linear
inhomogeneous equation for the first correction:
\begin{equation}
\begin{split}
\dot{\Pp}_i^{(1)}={}&\Om_i^{(1)}\times(\Pp_i^{(0)}-\epsc\ez)\\
&+\Om_i^{(0)}\times[\Pp_i^{(1)}+\epsc\PiR\cdot\Pp_j^{(0)}]\\
&-\Pp_i^{(1)}-\sigc\PiR\cdot\Pp_j^{(0)}
 -\epsc\PiR\cdot\dot{\Pp}_j^{(0)}.
\end{split}
\label{eq:Pfirst}
\end{equation}
The equations are solved in succession: first Eq.~\eqref{eq:Pzero} for
$\Pp_i^{(0)}$, then Eq.~\eqref{eq:Pfirst} for $\Pp_i^{(1)}$. The
first-order approximation to the dipole is the sum
$\Pp_i^{(0)}+\delta\Pp_i^{(1)}$. For it to be useful, the correction
$\delta\Pp_i^{(1)}$ must remain small compared with the part of the
leading solution being approximated. We examine this requirement first
for steady rotation and long times, and then near the critical field.

\section{Steady rotation and slow changes of the rotation axes}
\label{sec:slowrotation}
\subsection{Non-rotating state and isolated rotating family}
The labels `$\mathrm{nr}$' and `$\mathrm{rot}$' denote non-rotating and rotating
equilibria, respectively; superscripts $(0)$ and $(1)$ continue to denote
regular expansion coefficients. The isolated non-rotating state is
$\Pp_{\mathrm{nr}}^{(0)}=\sigc\ez$. At finite separation, setting
$\Om_i=\boldsymbol0$ and $\dot{\Pp}_i=\boldsymbol0$ in
Eq.~\eqref{eq:modelP} gives $\Pp_i=\sigc\boldsymbol E_i$. Subtracting the two field
equations shows that the dipoles for the two particles are the same for small $\delta$, yielding
\begin{align}
\Pp_{\mathrm{nr}}
 &=\sigc\ez-\delta\sigc^2\PiR\cdot\ez+\order(\delta^2),\label{eq:nr}\\
\Pp_{\mathrm{nr}}^{(0)}&=\sigc\ez,\qquad
\Pp_{\mathrm{nr}}^{(1)}=-\sigc^2\PiR\cdot\ez.
\label{eq:nrfirst}
\end{align}
In the non-rotating state, the dipole is collinear with its local
field, so the electric torque vanishes. Their common direction is
generally tilted relative to the imposed field.

Throughout the paper, transverse means perpendicular to the imposed field
$E_0\ez$, for example, $\Pp_{i,\perp}=(\I-\ez\ez)\cdot\Pp_i$.
For the general isolated-sphere dynamics, decompose
$\Pp^{(0)}=P_z^{(0)}\ez+\Pp_\perp^{(0)}$. Equation~\eqref{eq:Pzero} becomes
\begin{align}
\dot{\Pp}_\perp^{(0)}
 &=\left[\lambda_0-\frac{P_z^{(0)}-\sigc}{\Ma}\right]\Pp_\perp^{(0)},
 \label{eq:isolatedperp}\\
\dot P_z^{(0)}&=\Ma^{-1}|\Pp_\perp^{(0)}|^2-(P_z^{(0)}-\sigc),
 \label{eq:isolatedz}
\end{align}
where
\begin{equation}
\lambda_0=\frac{\Mac}{\Ma}-1
 =\left(\frac{E_0}{E_c^{(0)}}\right)^2-1.
\label{eq:lambda}
\end{equation}
A small transverse dipole perturbation about the non-rotating state
grows or decays at rate $\lambda_0$, whereas a perturbation along
the field decays at rate $1$. For $\lambda_0>0$, the non-rotating
state is unstable, and steady rotating states exist with arbitrary
azimuth:
\begin{equation}
\begin{split}
\Pp_{\mathrm{rot}}^{(0)}(\phi)
 &=(\epsc-\Ma)\ez
   +\Pperp^{(0)}\boldsymbol e_\rho(\phi), \\
\Pperp^{(0)}&=\Ma\sqrt{\lambda_0}.
 \label{eq:rot_p}
\end{split}
\end{equation}
Here $\Pperp^{(0)}$ is the steady value of $|\Pp_\perp^{(0)}|$
on the isolated rotating branch. In simulations the initial
perturbation selects its azimuth and rotation sense; in experiments
imperfections and fluctuations provide that perturbation. Interactions
can subsequently select a collective pattern. The transverse
basis vectors are
\begin{equation}
\boldsymbol e_\rho=(\cos\phi,\sin\phi,0),
\qquad
\boldsymbol e_\phi=\ez\times\boldsymbol e_\rho.
\label{eq:transversebasis}
\end{equation}
These are the radial and azimuthal unit vectors in the plane
perpendicular to the imposed field, with $\boldsymbol e_\rho$
pointing along the transverse dipole. The azimuthal angle $\phi$
therefore fixes the orientation of the rotation axis.
Substitution of Eq.~\eqref{eq:rot_p} into Eq.~\eqref{eq:Ozero} gives
$\Om_{\mathrm{rot}}^{(0)}=-\sqrt{\lambda_0}\,\boldsymbol e_\phi$.
The rotation axis is therefore perpendicular to the transverse dipole,
and the angular speed is $\sqrt{\lambda_0}$.

\subsection{Solvability of the first-order steady problem}
We now choose the leading dipoles to be steady rotating solutions
\eqref{eq:rot_p}, with prescribed azimuthal angles $\phi_i$. We use
\emph{phase} to mean this azimuthal angle throughout the analysis.
Since $\dot{\Pp}_j^{(0)}=\boldsymbol0$, the last term in
Eq.~\eqref{eq:Pfirst} vanishes for this choice of leading solution.
Substituting Eq.~\eqref{eq:Ofirst} into Eq.~\eqref{eq:Pfirst}
and collecting the terms proportional to $\Pp_i^{(1)}$ gives
\begin{equation}
\dot{\Pp}_i^{(1)}=\boldsymbol J_i\cdot\Pp_i^{(1)}+\boldsymbol G_i.
\label{eq:forced}
\end{equation}
Here $\boldsymbol J_i$ is the matrix obtained by linearizing the isolated
dipole equation about $\Pp_i^{(0)}$. The known vector $\boldsymbol G_i$
is the forcing caused by the other particle. To evaluate it, first define
the part of $\Om_i^{(1)}$ in Eq.~\eqref{eq:Ofirst} produced by particle $j$,
\begin{equation}
\Om_{i,\mathrm{int}}^{(1)}=-\Ma^{-1}
\left[\Pp_i^{(0)}\times(\PiR\cdot\Pp_j^{(0)})
+\tfrac12\PiR\cdot(\Pp_j^{(0)}\times\ez)\right].
\end{equation}
Substitution into Eq.~\eqref{eq:Pfirst} gives
\begin{equation}
\begin{split}
\boldsymbol G_i={}&\Om_{i,\mathrm{int}}^{(1)}
 \times(\Pp_i^{(0)}-\epsc\ez)\\
&+\epsc\Om_i^{(0)}\times(\PiR\cdot\Pp_j^{(0)})\\
&-\sigc\PiR\cdot\Pp_j^{(0)}.
\end{split}
\label{eq:G}
\end{equation}

Resolve the correction along $\boldsymbol e_{\rho i}$,
$\boldsymbol e_{\phi i}$, and $\ez$. These components change,
respectively, the transverse dipole magnitude, its direction, and the
dipole component along the imposed field. In this basis,
\begin{equation}
[\boldsymbol J_i]_{(\rho,\phi,z)}=
\begin{pmatrix}0&0&-\Pperp^{(0)}/\Ma\\0&0&0\\2\Pperp^{(0)}/\Ma&0&-1\end{pmatrix}.
\label{eq:J}
\end{equation}
The second row is zero: a small change of the isolated sphere's phase
produces another steady state, so there is no restoring term for that
change. The three component equations are therefore
\begin{equation}
\begin{split}
\dot P_{i,\rho}^{(1)}&=-\frac{\Pperp^{(0)}}{\Ma}P_{i,z}^{(1)}+G_{i,\rho}, \\
\dot P_{i,\phi}^{(1)}&=G_{i,\phi}, \\
\dot P_{i,z}^{(1)}&=\frac{2\Pperp^{(0)}}{\Ma}P_{i,\rho}^{(1)}-P_{i,z}^{(1)}+G_{i,z}.\label{eq:firstcomponents}
\end{split}
\end{equation}
For a time-independent first-order correction to steady rotation,
all three left-hand sides must vanish. The
second equation has a steady solution only if its forcing vanishes.
Evaluating the $\boldsymbol e_{\phi i}$ component of Eq.~\eqref{eq:G} gives
\begin{equation}
\begin{split}
G_{i,\phi}&=\frac{\Pperp^{(0)}}{2}
\left(\boldsymbol e_{\rho i}+\frac{\Pperp^{(0)}}{\Ma}\ez\right)
\cdot\PiR\cdot\boldsymbol e_{\phi j}\\
&=0,\qquad i=1,2.
\end{split}
\label{eq:phaseforcing}
\end{equation}
This is the \emph{solvability condition} for the first-order steady
problem. Equivalently, multiplying
$\boldsymbol J_i\Pp_i^{(1)}=-\boldsymbol G_i$ by the left null vector
$\boldsymbol e_{\phi i}$ requires the forcing to have zero component
along that vector, as in the solvability procedure described by
Nayfeh~\cite{Nayfeh1993}. When this condition holds, the first and third
equations give
\begin{align}
P_{i,z}^{(1)}&=\frac{\Ma}{\Pperp^{(0)}} G_{i,\rho},\label{eq:steadyz}\\
P_{i,\rho}^{(1)}&=\frac{\Ma}{2\Pperp^{(0)}}
\left(\frac{\Ma}{\Pperp^{(0)}} G_{i,\rho}-G_{i,z}\right).
\label{eq:steadyrho}
\end{align}
A constant $P_{i,\phi}^{(1)}$ merely changes the chosen phase by an
amount of order $\delta$; we set it to zero when specifying a steady state.

\subsection{Secular growth and a slow time for the phases}

For fixed azimuthal angles $\phi_1$ and $\phi_2$ of the leading
isolated-sphere solutions that do not satisfy
Eq.~\eqref{eq:phaseforcing}, the azimuthal component of
Eq.~\eqref{eq:firstcomponents} gives
\begin{equation}
P_{i,\phi}^{(1)}(t)
=P_{i,\phi}^{(1)}(0)+G_{i,\phi}t.
\label{eq:secular}
\end{equation}
The term proportional to $t$ is a \emph{secular term}: it eventually
makes the dipole correction $\delta P_{i,\phi}^{(1)}$ comparable
to the leading transverse magnitude $\Pperp^{(0)}$. It represents
a change of dipole direction. Indeed, a small phase change gives
\begin{equation}
\boldsymbol e_\rho(\phi_i+\Delta\phi_i)
=\boldsymbol e_\rho(\phi_i)
+\Delta\phi_i\boldsymbol e_{\phi i}
+\order(\Delta\phi_i^2),
\end{equation}
so
\begin{equation}
\Delta\phi_i\simeq
\frac{\delta P_{i,\phi}^{(1)}}{\Pperp^{(0)}}.
\end{equation}

To remove the secular growth, we continue to expand about the
isolated rotating family in Eq.~\eqref{eq:rot_p}, with
$\lambda_0>0$ fixed as $\delta\to0$. The leading transverse
dipole magnitude $\Pperp^{(0)}=\Ma\sqrt{\lambda_0}$ therefore
remains nonzero in this limit. We apply the method of multiple
scales~\cite{Nayfeh1993,BenderOrszag}, retaining the original
time $t$ and introducing the slow time $\tau=\delta t$.
The azimuthal angles of the leading rotating states are now
allowed to depend on $\tau$. We absorb this slow reorientation
into the leading term and write
\begin{equation}
\Pp_i
=\Pp_{\mathrm{rot}}^{(0)}(\phi_i(\tau))
+\delta\Pp_i^{(1)}(t,\tau)+\cdots.
\label{eq:outeransatz}
\end{equation}
Here $\Pp_i^{(1)}(t,\tau)$ denotes the remaining first-order
dipole correction, which is required to remain bounded in the
fast time $t$.

During the calculation, $t$ and $\tau$ are treated as independent
variables, with
\begin{equation}
\frac{\dd}{\dd t}
=\partial_t+\delta\partial_\tau.
\end{equation}
Substitution into Eqs.~\eqref{eq:model} gives, at order $\delta$,
\begin{equation}
\partial_t\Pp_i^{(1)}
=\boldsymbol J_i\cdot\Pp_i^{(1)}
+\boldsymbol G_i
-\Pperp^{(0)}\frac{\dd\phi_i}{\dd\tau}
 \boldsymbol e_{\phi i}.
\label{eq:outerforced}
\end{equation}
Its component along $\boldsymbol e_{\phi i}$ would grow
linearly in the fast time unless
\begin{equation}
\Pperp^{(0)}\frac{\dd\phi_i}{\dd\tau}
=G_{i,\phi}(\phi_1,\phi_2).
\label{eq:outerphase}
\end{equation}

Equation~\eqref{eq:outerphase} determines the slow evolution
of the rotation axes; its equilibria satisfy $G_{i,\phi}=0$.
We use it in the following subsection to determine the rotation
patterns and their stability away from onset. Section~\ref{sec:breakdown} determines how close to the isolated
threshold the interaction correction remains small relative
to the leading transverse dipole magnitude. Section~\ref{sec:multiple}
then develops an expansion about the critical non-rotating state.

\subsection{Phase selection away from onset}
\label{sec:phaseselection}

We now determine which rotating states are stable to small changes
in the directions of their rotation axes. We consider
$\lambda_0>0$ and $\delta/\lambda_0\ll1$, so that the interaction
correction remains small compared with the leading transverse
dipole magnitude. This ordering condition is derived in
Sec.~\ref{sec:breakdown}. In this regime, changes in magnitude
relax faster than the phases evolve, and the slow phase equation
determines the leading selection of the rotation axes.

Choose the line of centres in the $x$--$z$ plane,
$\widehat{\boldsymbol R}=(\sin\Theta,0,\cos\Theta)$.
Equation~\eqref{eq:outerphase} then becomes
\begin{equation}
\begin{split}
\frac{\dd\phi_i}{\dd\tau}=\frac12\bigl[
&\sin\phi_i\cos\phi_j\\
&-(\Pi_{xx}\cos\phi_i
+\sqrt{\lambda_0}\Pi_{zx})\sin\phi_j
\bigr],
\end{split}
\label{eq:phasecomponents}
\end{equation}
where $\Pi_{xx}=3\cos^2\Theta-2$ and
$\Pi_{zx}=-3\sin\Theta\cos\Theta$.
As established from Eqs.~\eqref{eq:rot_p} and \eqref{eq:Ozero},
$\Om_i^{(0)}=-\sqrt{\lambda_0}\,\boldsymbol e_{\phi i}$.
Thus equal phases give angular velocities in the same direction,
corresponding to co-rotation, whereas phases differing by $\pi$
give angular velocities in opposite directions, corresponding
to counter-rotation.

We first examine the counter-rotating branch
$(\phi_1,\phi_2)=(0,\pi)$; $(\pi,0)$ reverses both rotations. Its transverse dipoles point in opposite
directions along the $x$ axis, and its angular velocities point
in opposite directions along the $y$ axis. Substituting these
phases into Eq.~\eqref{eq:phasecomponents} makes both evolution
rates vanish for every pair angle. This branch therefore exists
throughout the parameter range considered.

To determine its stability, we linearize
Eq.~\eqref{eq:phasecomponents} with respect to the two phases.
The resulting phase Jacobian has trace $-1$ and determinant
\begin{equation}
\det\boldsymbol J_\phi
=\tfrac34\sin^2\Theta
\left[3(1+\lambda_0)\cos^2\Theta-1\right].
\label{eq:phasecounterdet}
\end{equation}
For $0<\Theta<\pi/2$, a positive determinant gives two eigenvalues
with negative real parts, whereas a negative determinant gives
one positive eigenvalue. The counter-rotating branch is therefore
stable for $\Theta<\Theta_*$ and unstable for $\Theta>\Theta_*$,
where
\begin{equation}
\cos^2\Theta_*=\frac{1}{3(1+\lambda_0)},
\qquad
\Theta_*=
\cos^{-1}\!\left(\frac{E_c^{(0)}}{\sqrt3E_0}\right).
\label{eq:outerselection}
\end{equation}
At $\Theta=0$, a common change of azimuth is neutral because
the geometry is axisymmetric about the imposed field.

We next examine co-rotating states. Setting
$\phi_1=\phi_2=\phi$ in Eq.~\eqref{eq:phasecomponents},
the equilibrium condition for co-rotating states factorizes as
\begin{equation}
0=\frac12\sin\phi
\left[(1-\Pi_{xx})\cos\phi
-\sqrt{\lambda_0}\Pi_{zx}\right].
\end{equation}
There are consequently two types of co-rotating solution.
The first has $\phi=0$ or $\pi$, so that both transverse dipoles
point along the same direction of the $x$ axis. These states
exist for every pair angle. However, their phase Jacobian has
trace $+1$, so at least one perturbation grows: they are unstable
throughout the regime considered.

The second co-rotating family satisfies
\begin{equation}
\phi_1=\phi_2=\phi,
\qquad
\cos\phi=-\sqrt{\lambda_0}\cot\Theta.
\label{eq:outercorotation}
\end{equation}
Here the common dipole direction changes with pair angle and
field strength. A real phase requires $|\cos\phi|\leq1$,
which gives the existence condition
\begin{equation}
\Theta\geq\Theta_{\mathrm{ex}},
\qquad
\Theta_{\mathrm{ex}}=\tan^{-1}\sqrt{\lambda_0}
=\cos^{-1}\!\left(\frac{E_c^{(0)}}{E_0}\right).
\end{equation}
For $\Theta>\Theta_{\mathrm{ex}}$, this family has
$\sin\phi\ne0$. At $\Theta=\Theta_{\mathrm{ex}}$,
$\cos\phi=-1$, so it meets the co-rotating state
$(\phi_1,\phi_2)=(\pi,\pi)$.
For the branch
in Eq.~\eqref{eq:outercorotation}, perturbations that change both
phases equally have growth rate $\nu_{\mathrm s}$, whereas
perturbations that increase one phase and decrease the other
have growth rate $\nu_{\mathrm a}$, with
\begin{align}
\nu_{\mathrm s}
&=-\frac32\sin^2\Theta\sin^2\phi,\\
\nu_{\mathrm a}
&=\frac12\left[3(1+\lambda_0)\cos^2\Theta-1\right].
\end{align}
These rates are measured on the slow time $\tau$.
Within the interior of the branch, $\nu_{\mathrm s}<0$.
The antisymmetric rate $\nu_{\mathrm a}$ is positive for
$\Theta<\Theta_*$ and negative for $\Theta>\Theta_*$.
This co-rotating branch therefore exists but is unstable for
$\Theta_{\mathrm{ex}}<\Theta<\Theta_*$, and becomes stable
for $\Theta>\Theta_*$.
At $\Theta=\pi/2$, its transverse dipoles lie along
$\boldsymbol e_y$ or $-\boldsymbol e_y$, and the corresponding
rotation axes lie along the $x$ axis.
A further branch with $\phi_2=-\phi_1$ can exist at oblique angles,
but is unstable there, as shown in Sec.~\hyperref[sec:outerphaseSM]{SII} of the SM.

\begin{figure}[!t]
\centering
\includegraphics[width=\columnwidth]
{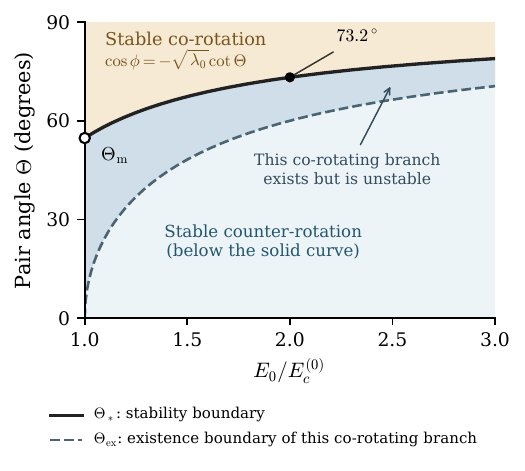}
\caption{Phase selection away from onset. The solid stability boundary
$\Theta_*$ separates stable counter-rotation, $(\phi_1,\phi_2)=(0,\pi)$,
below it from the stable co-rotating family of
Eq.~\eqref{eq:outercorotation} above it. That family exists above the
dashed boundary $\Theta_{\mathrm{ex}}$ and is unstable between the
curves. The co-rotating states $(0,0)$ and $(\pi,\pi)$ exist throughout
but are unstable. The filled circle marks $\Theta_*=73.2^\circ$ at
$E_0/E_c^{(0)}=2$. Both curves follow from the slow phase equations for
$R^{-3}\ll(E_0/E_c^{(0)})^2-1$, where perturbations to the rotation
speed relax faster than the rotation axes turn. The open circle marks
the limiting value of the solid curve, $\Theta_{\mathrm m}\simeq54.7^\circ$,
as the field approaches the isolated threshold within this regime.}
\label{fig:outerselection}
\end{figure}

As an illustration, consider a line of centres perpendicular to the
imposed field, $\Theta=\pi/2$. In this geometry the interaction tensor
is diagonal, making the effect on the two axis directions explicit. With
$\widehat{\boldsymbol R}=\boldsymbol e_x$, the ambient fluid
rotation at sphere $i$ produced by the other sphere's rotlet follows
from Eq.~\eqref{eq:modelO}:
\begin{equation}
-\frac{\delta}{2}\PiR\cdot\Om_j^{(0)}
=\delta\Omega_{j,x}^{(0)}\boldsymbol e_x
-\frac{\delta}{2}\Omega_{j,y}^{(0)}\boldsymbol e_y,
\end{equation}
where $\Omega_{j,z}^{(0)}=0$. The rotlet flow therefore reinforces
the angular-velocity component along the line of centres and opposes
the component along $y$. In the states $(0,0)$ and $(\pi,\pi)$,
both spheres rotate about $y$. A small common tilt of their axes
towards $x$ introduces angular-velocity components that the other
sphere's rotlet flow reinforces, so the tilt grows. In the states
with $\phi=\pi/2$ or $3\pi/2$, both spheres rotate about the line of
centres. A small common tilt towards $y$ introduces components
that the rotlet flow opposes, restoring the original orientation.

For perturbations that tilt both axes together, the equal-phase equation
at $\Theta=\pi/2$ is
\begin{equation}
\frac{\dd\phi}{\dd\tau}=\frac34\sin(2\phi).
\end{equation}
Its linear growth rate is $+3/2$ at $\phi=0$ or $\pi$, and
$-3/2$ at $\phi=\pi/2$ or $3\pi/2$. For the latter states,
the opposite-phase perturbation also decays, with rate
$\nu_{\mathrm a}=-1/2$. Thus both phase perturbations decay
when the two spheres co-rotate along the line of centres.

Figure~\ref{fig:outerselection} distinguishes the existence
boundary $\Theta_{\mathrm{ex}}$ from the stability boundary
$\Theta_*$. For example, at $E_0/E_c^{(0)}=2$,
$\lambda_0=3$ and $\Theta_{\mathrm{ex}}=60^\circ$, whereas
$\Theta_*=73.2^\circ$. Between these angles, both the
counter-rotating branch and the co-rotating branch in
Eq.~\eqref{eq:outercorotation} exist, but only counter-rotation
is stable. Above $\Theta_*$, counter-rotation is unstable and
the latter co-rotating branch is stable. The simpler co-rotating
states with $\phi=0$ or $\pi$ remain unstable on both sides
of this boundary.

At $\Theta=\Theta_*$, the leading phase equations possess a
continuous family of equilibria connecting the counter-rotating
and co-rotating states. They therefore do not select a unique
rotation pattern exactly on the boundary. Higher-order terms
are needed to resolve the branch selection there. The phase
growth rates and this family of equilibria are derived in
Sec.~\hyperref[sec:outerphaseSM]{SII} of the SM.

The condition $\delta/\lambda_0\ll1$ is equivalent to
$R^{-3}\ll(E_0/E_c^{(0)})^2-1$, which fails at fixed separation as
the isolated threshold is approached. The open circle at
$\Theta_{\mathrm m}=\cos^{-1}(1/\sqrt3)\simeq54.7^\circ$
is the limit of $\Theta_*$ within the overlap regime
$\delta\ll\lambda_0\ll1$. At $\lambda_0=\order(\delta)$,
magnitude and direction evolve together (Sec.~\ref{sec:multiple}).

\subsection{Long-time behavior of the regular approximation}
\begin{figure*}
\centering
\includegraphics[width=.98\textwidth]{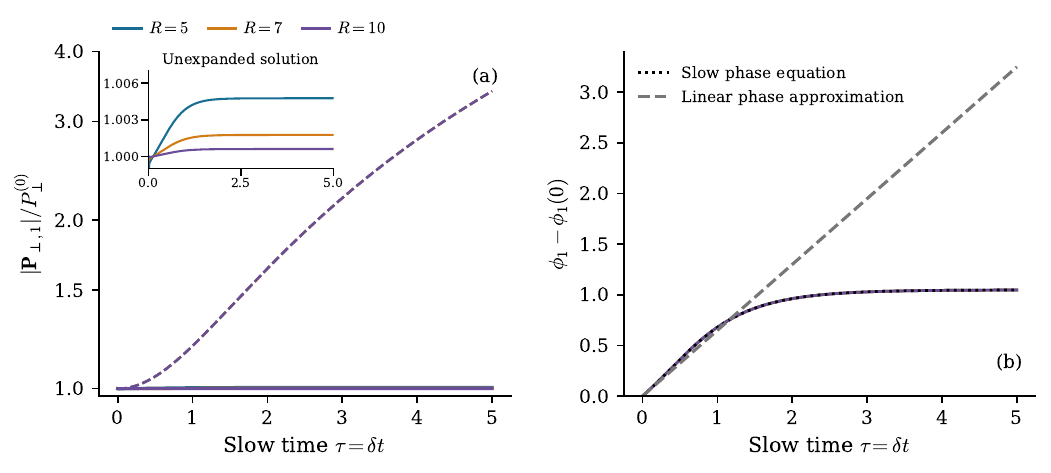}
\caption{Loss of uniform validity of the regular expansion at long times for
$\widehat{\boldsymbol R}=\boldsymbol e_x$, $\lambda_0=1$, and
$\phi_1(0)=\phi_2(0)=\pi/6$. (a) Transverse dipole magnitude from the unexpanded
model (solid) and the sum $\Pp^{(0)}+\delta\Pp^{(1)}$ (dashed),
normalized by $\Pperp^{(0)}$. The main vertical axis is logarithmic; the inset
shows the bounded unexpanded solutions on an expanded linear scale.
The dashed magnitude curves nearly overlap on this scale. (b) Phase change from the unexpanded model (colored curves), the slow
phase equation (black dotted), and the linear phase approximation
$\delta P_{1,\phi}^{(1)}/\Pperp^{(0)}$ (gray dashed).
The unexpanded curves lie close to the dotted slow-phase curve and
approach a phase change of $\pi/3$,
corresponding to a final phase of $\pi/2$.
}
\label{fig:secular}
\end{figure*}
We first return to the regular expansion with fixed azimuthal angles
$\phi_1$ and $\phi_2$ of the isolated rotating solutions. Here $\Pp_i^{(1)}$ is the coefficient in
Eq.~\eqref{eq:regularP}, before slow reorientation is absorbed
into the leading term as in Eq.~\eqref{eq:outeransatz}.
We integrate Eqs.~\eqref{eq:Pzero} and \eqref{eq:Pfirst}, form the sum
$\Pp_i^{(0)}+\delta\Pp_i^{(1)}$, and compare this approximate dipole
with a direct integration of the unexpanded model. Here
$\epsc=-0.1097$, $\sigc=-0.5$, and $\lambda_0=1$, so
$E_0/E_c^{(0)}=\sqrt2$. The line of centres is perpendicular to the
field, $\widehat{\boldsymbol R}=\boldsymbol e_x$, and the leading
dipoles initially have $\phi_1=\phi_2=\pi/6$. We set
$\Pp_i^{(1)}(0)=\boldsymbol0$ and use the same initial dipoles for
the unexpanded model. The leading dipoles are steady isolated solutions.

For these fixed reference phases, Eq.~\eqref{eq:secular} gives a
phase change that grows at the slow-time rate
\begin{equation}
\frac{G_{i,\phi}}{\Pperp^{(0)}}=\frac34\sin(2\phi_i)=\frac{3\sqrt3}{8}.
\end{equation}
The actual dipole correction grows at rate $\delta G_{i,\phi}$. As the secular term grows, the sum $\Pp_i^{(0)}+\delta\Pp_i^{(1)}$ also acquires an
artificially increasing magnitude [Fig.~\ref{fig:secular}(a)]. The
unexpanded transverse magnitude approaches a constant value
[Fig.~\ref{fig:secular}(a), inset]. In panel (b), the phase
change $\phi_1-\phi_1(0)$ approaches $\pi/3$; since
$\phi_1(0)=\pi/6$, the phase itself approaches $\pi/2$.
Equation~\eqref{eq:outerphase} reduces here to
$\dd\phi/\dd\tau=(3/4)\sin(2\phi)$ and describes that gradual change
of direction [Fig.~\ref{fig:secular}(b)].

\begin{figure}
\centering
\includegraphics[width=\columnwidth]{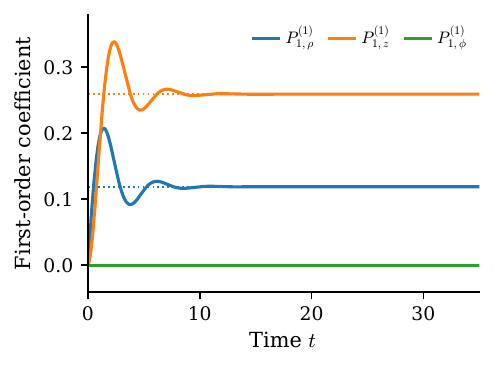}
\caption{Bounded first-order coefficients when the phases satisfy the
solvability condition: $\phi_1=\phi_2=\pi/2$ and
$\widehat{\boldsymbol R}=\boldsymbol e_x$. Other parameters are as in
Fig.~\ref{fig:secular}. Dotted lines are the steady values from
Eqs.~\eqref{eq:steadyz} and \eqref{eq:steadyrho}; the phase component
vanishes. The dipole correction is $\delta$ times each plotted coefficient.}
\label{fig:locked}
\end{figure}

Next choose $\phi_1=\phi_2=\pi/2$, keeping the other parameters fixed.
Now $G_{i,\phi}=0$, so the solvability condition is satisfied and there
is no secular phase term. The first-order coefficients approach
$P_{i,\rho}^{(1)}=0.1190$ and $P_{i,z}^{(1)}=0.2593$
(Fig.~\ref{fig:locked}). These steady coefficients follow from Eqs.~\eqref{eq:steadyz} and
\eqref{eq:steadyrho}.
Thus its steady transverse magnitude is
$\Pperp^{(0)}+0.1190\delta+\order(\delta^2)$, and its longitudinal
dipole is $\epsc-\Ma+0.2593\delta+\order(\delta^2)$.

\section{Failure of the regular expansion near onset}
\label{sec:breakdown}

\begin{figure*}
\centering
\includegraphics[width=.98\textwidth]{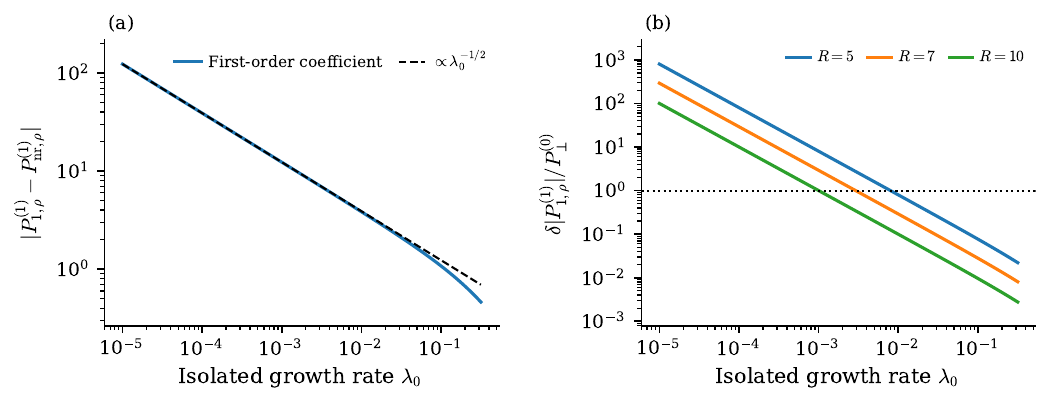}
\caption{Breakdown of the regular expansion for the transverse dipole
magnitude near onset for $\widehat{\boldsymbol R}=\boldsymbol e_x$ and
$\phi_1=\phi_2=\pi/2$. (a) Magnitude of the first-order radial dipole
coefficient relative to the non-rotating state,
$\lvert P_{1,\rho}^{(1)}-P_{\mathrm{nr},\rho}^{(1)}\rvert$,
and its asymptote $\Mac/\sqrt{\lambda_0}$.
(b) Magnitude of the first-order correction relative to the isolated
transverse dipole magnitude, $\delta\lvert P_{1,\rho}^{(1)}\rvert/\Pperp^{(0)}$.
The horizontal line marks a ratio of one. We vary $\lambda_0$ by
changing the applied field while keeping the material parameters of
Fig.~\ref{fig:locked} fixed.}
\label{fig:threshold}
\end{figure*}
We now examine the steady first-order magnitude correction as
the isolated critical field is approached, $\lambda_0\to0^+$.
Choose phases satisfying Eq.~\eqref{eq:phaseforcing}, so that
$G_{i,\phi}=0$ and there is no secular phase forcing. The question
is whether the transverse correction $\delta P_{i,\rho}^{(1)}$
remains small compared with the isolated transverse magnitude
$\Pperp^{(0)}$.

We use the first and third component equations in
Eq.~\eqref{eq:firstcomponents}. Setting their time derivatives
to zero gives
\begin{equation}
\begin{split}
0&=-\frac{\Pperp^{(0)}}{\Ma}P_{i,z}^{(1)}+G_{i,\rho},\\
0&=\frac{2\Pperp^{(0)}}{\Ma}P_{i,\rho}^{(1)}
   -P_{i,z}^{(1)}+G_{i,z}.
\end{split}
\label{eq:steadycomponents_breakdown}
\end{equation}
These are the two linear equations previously solved to obtain
Eqs.~\eqref{eq:steadyz} and \eqref{eq:steadyrho}. Their coefficient
matrix is the $(\rho,z)$ block of the Jacobian in
Eq.~\eqref{eq:J}, with determinant
\begin{equation}
\det[\boldsymbol J_i]_{(\rho,z)}
=\frac{2(\Pperp^{(0)})^2}{\Ma^2}=2\lambda_0.
\label{eq:det}
\end{equation}
The homogeneous perturbations governed by this block have growth
rates satisfying $s^2+s+2\lambda_0=0$. As $\lambda_0\to0^+$,
one tends to $-1$ and the other behaves as $-2\lambda_0$.
The relaxation of the rotation magnitude therefore becomes slow
near onset, allowing a weak interaction to produce a large
magnitude correction.

To determine the size of that correction, we must also evaluate
the forcing. The first equation in
Eq.~\eqref{eq:steadycomponents_breakdown} gives
\begin{equation}
P_{i,z}^{(1)}=\frac{\Ma}{\Pperp^{(0)}}G_{i,\rho},
\end{equation}
which is Eq.~\eqref{eq:steadyz}. Substituting this expression into
the second equation and using
$(\Pperp^{(0)})^2=\Ma^2\lambda_0$ gives
\begin{equation}
2\lambda_0 P_{i,\rho}^{(1)}
=G_{i,\rho}-\frac{\Pperp^{(0)}}{\Ma}G_{i,z}.
\label{eq:effectiveforcing}
\end{equation}
This is Eq.~\eqref{eq:steadyrho} rearranged to display the small
factor multiplying the unknown magnitude correction.

The right-hand side also becomes small near onset. To evaluate
it, substitute the isolated rotating dipoles from
Eq.~\eqref{eq:rot_p} into the forcing in Eq.~\eqref{eq:G}.
Using Eq.~\eqref{eq:lambda} to write
\begin{equation}
\Ma=\frac{\Mac}{1+\lambda_0},
\qquad
\Pperp^{(0)}=\Ma\sqrt{\lambda_0},
\end{equation}
and expanding at fixed pair orientation gives
\begin{equation}
\begin{split}
&G_{i,\rho}-\frac{\Pperp^{(0)}}{\Ma}G_{i,z}\\
&\quad=\Pperp^{(0)}\bigl[-2\sigc\ez\cdot\PiR\cdot\ez
 -\tfrac12\boldsymbol e_{\phi i}\cdot\PiR\cdot\boldsymbol e_{\phi j}\bigr]\\
&\qquad+\order\!\left((\Pperp^{(0)})^2\right).
\end{split}
\label{eq:radialcoefficient}
\end{equation}
Dividing by $2\lambda_0=2(\Pperp^{(0)})^2/\Ma^2$ therefore yields
\begin{equation}
\begin{split}
P_{i,\rho}^{(1)}={}&\frac{\Ma^2}{2\Pperp^{(0)}}
\bigl[-2\sigc\ez\cdot\PiR\cdot\ez\\
&\hspace{12mm}
-\tfrac12\boldsymbol e_{\phi i}\cdot\PiR\cdot\boldsymbol e_{\phi j}\bigr]
+\order(1).
\end{split}
\label{eq:radiallimit}
\end{equation}
When the coefficient in brackets approaches a nonzero value,
$P_{i,\rho}^{(1)}=\order(\lambda_0^{-1/2})$.
The same expansion of Eq.~\eqref{eq:G} gives
$G_{i,\rho}=\order(\Pperp^{(0)})$, so
Eq.~\eqref{eq:steadyz} shows that the longitudinal correction
$P_{i,z}^{(1)}$ remains bounded.

For these generic configurations, the transverse correction
relative to the isolated transverse dipole has size
\begin{equation}
\frac{\delta|P_{i,\rho}^{(1)}|}{\Pperp^{(0)}}
=\order\!\left(\frac{\delta}{\lambda_0}\right).
\label{eq:orderratio}
\end{equation}
The regular expansion therefore requires $\lambda_0\gg\delta$.
When $\lambda_0=\order(\delta)$, the first-order transverse
correction is as large as the leading transverse dipole, and
the expansion about the isolated rotating state loses its
ordering in those components. Since
$\Pperp^{(0)}=\Ma\sqrt{\lambda_0}$, the corresponding balance is
\begin{equation}
\Pperp^{(0)}=\order(\delta^{1/2}).
\label{eq:distinguished}
\end{equation}
If the coefficient in brackets in Eq.~\eqref{eq:radiallimit}
vanishes at onset, the leading singular term is absent and
further terms must be examined for that configuration.

In Fig.~\ref{fig:threshold}(a), the blue curve is the first-order coefficient
from the steady correction equations, and the dashed curve is its
$\lambda_0^{-1/2}$ asymptote. Their separation at larger $\lambda_0$
reflects the higher-order terms.
To illustrate this breakdown, we use the same pair configuration
as in Fig.~\ref{fig:locked}, now varying $\lambda_0$ towards
zero. The line of centres is along $\boldsymbol e_x$, and both
transverse dipoles are along $\boldsymbol e_y$, giving
$\PiR=\operatorname{diag}(-2,1,1)$ and
$\boldsymbol e_{\phi i}=\boldsymbol e_{\phi j}=-\boldsymbol e_x$.
For this geometry, the non-rotating correction in
Eq.~\eqref{eq:nrfirst} has no $\boldsymbol e_y$ component, so
the radial correction relative to the non-rotating state is
simply $P_{i,\rho}^{(1)}$.
With $\sigc=-1/2$, the square-bracketed factor in
Eq.~\eqref{eq:radiallimit} equals $2$, confirming that the
leading singular term does not cancel.
Figure~\ref{fig:threshold}(a) shows the resulting growth of
$P_{i,\rho}^{(1)}$ as onset is approached, together with its
$\lambda_0^{-1/2}$ asymptote.

The consequence for the regular expansion is shown more
directly in Fig.~\ref{fig:threshold}(b), which plots the
actual correction relative to the leading transverse dipole,
$\delta|P_{i,\rho}^{(1)}|/\Pperp^{(0)}$.
At fixed separation, this ratio increases as $\lambda_0$
decreases. For $\lambda_0\gg\delta$, the correction remains
small and the magnitude expansion is ordered.
When $\lambda_0=\order(\delta)$, the correction becomes
comparable to the leading term and the regular expansion
fails. The horizontal line marks equal sizes of these two
terms.

On the isolated rotating branch, the transverse dipole
magnitude and angular speed vanish as $\lambda_0\to0^+$,
while the longitudinal dipole tends to $\sigc$.
Interactions must therefore enter the leading balance
determining the small transverse dipole near onset.
The next section constructs this balance by expanding about
the critical non-rotating state, $\Pp=\sigc\ez$, with both
the magnitude and direction of the transverse dipole evolving
on the interaction time.

\section{Near-onset expansion about the critical non-rotating state}
\label{sec:multiple}
\subsection{Detuning, time scales, and expansion}
Section~\ref{sec:breakdown} identifies the balance
$\lambda_0=\order(\delta)$ for which the regular rotating-state
expansion fails. We therefore return to the governing
Eqs.~\eqref{eq:model} and construct an expansion about the isolated
non-rotating state at its critical field. The small transverse dipole
will be determined by the interaction and the departure from that field.

Using the isolated critical Mason number $\Mac$ defined in
Eq.~\eqref{eq:Mac}, introduce a small parameter
$\eta$ and a detuning parameter $\rho$ by
\begin{equation}
\eta = \delta^{1/2},\qquad \rho =\frac{\lambda_0}{\eta^2},\qquad
\Ma^{-1}=\frac{1+\rho\eta^2}{\Mac}.
\label{eq:detuning}
\end{equation}
Here $\rho$ measures the squared field's departure from isolated onset
relative to the interaction strength:
$\rho=[(E_0/E_c^{(0)})^2-1]/\eta^2$.
It remains of order one as $\eta\to0$ and can have either sign.
Positive $\rho$ corresponds to a field above the isolated threshold;
negative $\rho$ corresponds to a field below it. This use of a detuning parameter follows Ref.~\cite{Nayfeh1993}.

Introduce the time scales
\begin{equation}
\begin{split}
T_0&=t,\qquad T_1=\eta t,\qquad T_2=\eta^2t,\\
\frac{\dd}{\dd t}&=D_0+\eta D_1+\eta^2D_2,
\qquad D_n=\frac{\partial}{\partial T_n}.
\label{eq:times}
\end{split}
\end{equation}
We initially allow dependence on all three scales. The solvability
conditions will determine which slow scales are needed. The expansion for the dipole is
\begin{equation}
\Pp_i=\sigc\ez+\eta\Pp_i^{[1]}+
\eta^2\Pp_i^{[2]}+\eta^3\Pp_i^{[3]}+\order(\eta^4),
\label{eq:msansatz}
\end{equation}
where each coefficient may depend on $(T_0,T_1,T_2)$. Square brackets
distinguish the coefficients of powers of $\eta$ from the round-bracket
superscripts in the expansion in $\delta$, Eq.~\eqref{eq:regularP}.
In particular, $\Pp_i^{[2]}$ multiplies $\eta^2=\delta$, whereas
$\Pp_i^{(1)}$ is the first correction about an isolated rotating state.
The leading transverse dipole is of order $\eta$, as required by
Eq.~\eqref{eq:distinguished}.

We substitute Eqs.~\eqref{eq:detuning}--\eqref{eq:msansatz} into
each of Eqs.~\eqref{eq:model}. The local-field equation immediately gives
\begin{equation}
\boldsymbol E_i=\ez-\eta^2\sigc\PiR\cdot\ez
-\eta^3\PiR\cdot\Pp_j^{[1]}+\order(\eta^4).
\label{eq:msfield}
\end{equation}
We also expand the angular velocity as
$\Om_i=\eta\boldsymbol\omega_i^{[1]}+
\eta^2\boldsymbol\omega_i^{[2]}+
\eta^3\boldsymbol\omega_i^{[3]}+\cdots$.
At each order the torque equation determines its coefficient, which
is then substituted into the dipole equation. Equating like powers of
$\eta$ gives the following successive problems.

\subsection{Order \texorpdfstring{$\eta$}{eta}: the leading transverse dipole}
The torque and dipole equations give
\begin{equation}
\boldsymbol\omega_i^{[1]}=\Mac^{-1}\Pp_i^{[1]}\times\ez,
\qquad D_0\Pp_i^{[1]}=-(\Pp_i^{[1]}\cdot\ez)\ez.
\label{eq:fast}
\end{equation}
Hence
\begin{equation}
\Pp_i^{[1]}=\boldsymbol a_i(T_1,T_2)
+c_i(T_1,T_2)e^{-T_0}\ez,\qquad \boldsymbol a_i\cdot\ez=0.
\end{equation}
The longitudinal perturbation decays on the charge-relaxation
time. After this initial transient,
$\Pp_i^{[1]}=\boldsymbol a_i(T_1,T_2)$, so
$\eta\boldsymbol a_i$ is the leading transverse dipole.
Our aim is to derive coupled evolution equations for
$\boldsymbol a_1$ and $\boldsymbol a_2$, determining how
detuning and interactions affect both the magnitude and
direction of each transverse dipole.
Their dependence on the slow times is undetermined at this
order and will follow from the solvability conditions at
subsequent orders.

\subsection{Order \texorpdfstring{$\eta^2$}{eta squared}: the first solvability condition}
At this order the torque balance gives
\begin{equation}
\boldsymbol\omega_i^{[2]}=\Mac^{-1}
\left[\Pp_i^{[2]}+\sigc^2\PiR\cdot\ez\right]\times\ez.
\label{eq:msomega2}
\end{equation}
Using this expression in the dipole equation gives
\begin{equation}
\begin{split}
D_0\Pp_i^{[2]}+D_1\boldsymbol a_i
={}&\left[\frac{|\boldsymbol a_i|^2}{\Mac}
-\Pp_i^{[2]}\cdot\ez\right.\\
&\left.\hspace{7mm}-\sigc^2\ez\cdot\PiR\cdot\ez\right]\ez.
\label{eq:msorder2}
\end{split}
\end{equation}
The right-hand side has no transverse component. Thus
\begin{equation}
D_0\Pp_{i,\perp}^{[2]}=-D_1\boldsymbol a_i.
\label{eq:intermediatefast}
\end{equation}
Its solution contains $-T_0D_1\boldsymbol a_i$, which is secular.
Requiring a bounded correction gives the solvability condition
\begin{equation}
D_1\boldsymbol a_i=\boldsymbol0,
\qquad \boldsymbol a_i=\boldsymbol a_i(T_2).
\label{eq:nointermediate}
\end{equation}
The leading transverse amplitude $\boldsymbol a_i$ therefore
does not evolve on $T_1=\eta t$. Its evolution on $T_2$
remains to be determined.

The longitudinal part of Eq.~\eqref{eq:msorder2} relaxes to a value
determined by $|\boldsymbol a_i|^2$.
Equation~\eqref{eq:intermediatefast} determines the transverse part of
$\Pp_i^{[2]}$ only up to a vector independent of $T_0$.
Because the leading transverse dipole is $\eta\boldsymbol a_i$, an
additional contribution at order $\eta^2$ can be absorbed by an
order-$\eta$ adjustment of $\boldsymbol a_i$. We absorb all of this
contribution except the known tilt of the non-rotating state, so that
$\boldsymbol a_i$ measures the transverse dipole relative to that tilted
state. With this choice,
\begin{equation}
\Pp_i^{[2]}=-\sigc^2\PiR\cdot\ez
+\frac{|\boldsymbol a_i|^2}{\Mac}\ez,
\qquad \boldsymbol\omega_i^{[2]}=\boldsymbol0.
\label{eq:slaving}
\end{equation}
The first term is the electric interaction correction to the
non-rotating dipole found in Eq.~\eqref{eq:nrfirst}, with
$\delta=\eta^2$.
The second term is a longitudinal dipole correction generated
by convection of surface charge by the particle's own rotation.
This contribution is already present for an isolated rotating
sphere and vanishes on the non-rotating branch,
$\boldsymbol a_i=\boldsymbol0$.

\subsection{Order \texorpdfstring{$\eta^3$}{eta cubed}: evolution on the slow time}
The torque coefficient at this order is
\begin{equation}
\begin{split}
\boldsymbol\omega_i^{[3]}=\Mac^{-1}\bigl[&
\Pp_i^{[3]}\times\ez+\rho\boldsymbol a_i\times\ez
-\sigc\boldsymbol a_i\times(\PiR\cdot\ez)\\
&-\sigc\ez\times(\PiR\cdot\boldsymbol a_j)
-\tfrac12\PiR\cdot(\boldsymbol a_j\times\ez)\bigr].
\label{eq:msomega3}
\end{split}
\end{equation}
Substitute this expression and Eq.~\eqref{eq:slaving} into the
order-$\eta^3$ dipole equation. Its transverse part is
\begin{equation}
D_0\Pp_{i,\perp}^{[3]}=\boldsymbol{\mathcal F}_i
-D_2\boldsymbol a_i,
\label{eq:mssecular}
\end{equation}
where the known forcing is
\begin{equation}
\begin{split}
\boldsymbol{\mathcal F}_i={}&
 [\rho-2\sigc(\ez\cdot\PiR\cdot\ez)]\boldsymbol a_i
 -\frac{|\boldsymbol a_i|^2}{\Mac^2}\boldsymbol a_i\\
&+\frac12\ez\times[\PiR\cdot(\ez\times\boldsymbol a_j)].
\label{eq:msforcing}
\end{split}
\end{equation}

The right-hand side of Eq.~\eqref{eq:mssecular} is independent
of $T_0$. A nonzero difference
$\boldsymbol{\mathcal F}_i-D_2\boldsymbol a_i$ would therefore
produce a term proportional to $T_0$ in
$\Pp_{i,\perp}^{[3]}$. Eliminating this secular term gives the
solvability condition
$D_2\boldsymbol a_i=\boldsymbol{\mathcal F}_i$, or explicitly,
\begin{equation}
\begin{split}
\frac{\dd\boldsymbol a_i}{\dd T_2}={}&
 [\rho-2\sigc(\ez\cdot\PiR\cdot\ez)]\boldsymbol a_i
 -\frac{|\boldsymbol a_i|^2}{\Mac^2}\boldsymbol a_i\\
&+\frac12\ez\times[\PiR\cdot(\ez\times\boldsymbol a_j)].
\label{eq:amplitude}
\end{split}
\end{equation}
Equation~\eqref{eq:amplitude} is the central result of the near-onset analysis.
It determines both the magnitude and direction of each
leading transverse dipole, $\eta\boldsymbol a_i$, on the
interaction time $T_2=\eta^2t=\delta t$. The term $\rho\boldsymbol a_i$ describes growth or decay
associated with the departure of the applied field from the
isolated critical field. The electric interaction contributes
$-2\sigc(\ez\cdot\PiR\cdot\ez)\boldsymbol a_i$, through the
change in the local field and the associated non-rotating
polarization. Its coefficient depends on the pair angle,
but multiplies both transverse components by the same factor.
At this order, it shifts onset equally for every azimuthal direction.

The negative cubic term comes from the longitudinal dipole
correction in Eq.~\eqref{eq:slaving}, generated by convection
of surface charge by each particle's own rotation.
This local nonlinear feedback limits the growth of the
transverse dipole and is already present for an isolated
rotating sphere. Like the detuning and electric interaction
terms, it acts parallel to $\boldsymbol a_i$, changing its
magnitude without directly changing its direction.

The final term is the hydrodynamic interaction with the other
particle. It depends on $\boldsymbol a_j$ and retains the
directional dependence of the interaction tensor $\PiR$.
It can modify the growth of $\boldsymbol a_i$ and also have
a component perpendicular to it, allowing the dipole to
reorient. This coupling distinguishes between different
relative orientations of the two dipoles and different
rotation-axis directions. Thus the electric interaction
shifts the growth rates equally for all transverse modes,
whereas the hydrodynamic interaction produces the differences
between them and selects the preferred collective rotation pattern.

Each $\boldsymbol a_i$ has two scalar components perpendicular to the
field. Equation~\eqref{eq:amplitude} therefore consists of four coupled
scalar ordinary differential equations, one for each of
$X_1,Y_1,X_2,Y_2$ when
$\boldsymbol a_i=X_i\boldsymbol e_x+Y_i\boldsymbol e_y$.
Both the magnitude and direction of each transverse dipole can vary.

The local-field derivative in Eq.~\eqref{eq:modelP} contributes at order
$\eta^3$ during the initial transient through $D_0\Pp_j^{[1]}$.
Afterwards, Eq.~\eqref{eq:nointermediate} gives
$\epsc\dot{\boldsymbol E}_i=-\epsc\eta^5\PiR\cdot
\dd\boldsymbol a_j/\dd T_2+\cdots$, which is beyond the order retained in
Eq.~\eqref{eq:amplitude}.

\subsection{Connection with the phase dynamics away from onset}
For a nonzero amplitude, write
$\boldsymbol a_i=|\boldsymbol a_i|\boldsymbol e_{\rho i}$.
Projection of Eq.~\eqref{eq:amplitude} along $\boldsymbol e_{\phi i}$ gives
\begin{equation}
|\boldsymbol a_i|\frac{\dd\phi_i}{\dd T_2}
=\frac{|\boldsymbol a_j|}{2}
\boldsymbol e_{\rho i}\cdot\PiR\cdot\boldsymbol e_{\phi j}.
\label{eq:msphase}
\end{equation}
For equal leading magnitudes, the constant-phase condition agrees with
the $\Pperp^{(0)}\to0^+$ limit of Eq.~\eqref{eq:phaseforcing}.
The near-onset equations thus recover the limiting phase condition
while also determining the magnitude, which can no longer be prescribed
by the isolated-sphere solution.

Without interactions, Eq.~\eqref{eq:amplitude} reduces to
$\dd\boldsymbol a_i/\dd T_2=
\rho\boldsymbol a_i
-|\boldsymbol a_i|^2\boldsymbol a_i/\Mac^2$.
For $\rho>0$, its nonzero steady magnitude gives
\begin{equation}
|\Pp_{i,\perp}|=\Mac\sqrt{\lambda_0}
+\order(\lambda_0^{3/2}),
\end{equation}
in agreement with $\Pperp^{(0)}=\Ma\sqrt{\lambda_0}$, because
\begin{equation}
\Ma=\frac{\Mac}{1+\lambda_0},\qquad
\Ma\sqrt{\lambda_0}=\Mac\sqrt{\lambda_0}
+\order(\lambda_0^{3/2}).
\label{eq:Macmatching}
\end{equation}

\section{Four rotation patterns near onset}
\label{sec:modes}
\subsection{Linear growth rates and the four modes}
Choose the line of centres in the $x$--$z$ plane,
$\widehat{\boldsymbol R}=\sin\Theta\boldsymbol e_x+
\cos\Theta\ez$. For $\Theta\ne0$, $\boldsymbol e_x$ lies in the
plane containing the line of centres and the imposed field, and
$\boldsymbol e_y$ is perpendicular to that plane. Substituting
$\boldsymbol a_i=X_i\boldsymbol e_x+Y_i\boldsymbol e_y$ into
Eq.~\eqref{eq:amplitude} gives
\begin{align}
\frac{\dd X_i}{\dd T_2}&=gX_i-\tfrac12X_j
 -\frac{X_i^2+Y_i^2}{\Mac^2}X_i,\label{eq:Xamp}\\
\frac{\dd Y_i}{\dd T_2}&=gY_i+hY_j
 -\frac{X_i^2+Y_i^2}{\Mac^2}Y_i,\label{eq:Yamp}
\end{align}
where $j=3-i$ and
\begin{equation}
g=\rho-2\sigc(1-3\cos^2\Theta),\qquad
h=\tfrac12(2-3\cos^2\Theta).
\label{eq:gh}
\end{equation}
Here $g$ combines the detuning with the common electric shift.
The coupling coefficients $-1/2$ and $h$ come from the rotlet
interaction in Eq.~\eqref{eq:amplitude}.
To find the onset of rotation, linearize about
$X_1=Y_1=X_2=Y_2=0$ by omitting the cubic terms. The resulting equations are
\begin{equation}
\begin{split}
\frac{\dd}{\dd T_2}\begin{pmatrix}X_1\\X_2\end{pmatrix}
 &=\begin{pmatrix}g&-1/2\\-1/2&g\end{pmatrix}
 \begin{pmatrix}X_1\\X_2\end{pmatrix},\\
\frac{\dd}{\dd T_2}\begin{pmatrix}Y_1\\Y_2\end{pmatrix}
 &=\begin{pmatrix}g&h\\h&g\end{pmatrix}
 \begin{pmatrix}Y_1\\Y_2\end{pmatrix}.
\label{eq:linearcomponents}
\end{split}
\end{equation}
Exchanging the two identical particles leaves these linearized equations
unchanged. Perturbations that are equal on the two particles and those
that are opposite therefore evolve independently, as expected from
particle-exchange symmetry~\cite{Hoyle2006}. Adding and subtracting
the equations implements this decomposition. Define
\begin{equation}
X_\pm=\frac{X_1\pm X_2}{\sqrt2},\qquad
Y_\pm=\frac{Y_1\pm Y_2}{\sqrt2}.
\label{eq:paircoords}
\end{equation}
Each coordinate obeys $\dd Z_k/\dd T_2=\Lambda_k Z_k$, with
$Z_k=X_+,X_-,Y_+,Y_-$ and $\Lambda_k=\rho+m_k$, where
\begin{align}
m_{X_\pm}&=-2\sigc(1-3\cos^2\Theta)\mp\tfrac12,\label{eq:mX}\\
m_{Y_\pm}&=-2\sigc(1-3\cos^2\Theta)
 \pm\tfrac12(2-3\cos^2\Theta).\label{eq:mY}
\end{align}
The upper signs give $X_+$ and $Y_+$, and the lower signs give
$X_-$ and $Y_-$.
Thus $Z_k(T_2)=Z_k(0)e^{\Lambda_kT_2}$: the disturbance decays for
$\Lambda_k<0$ and grows for $\Lambda_k>0$.

\begin{figure*}
\centering
\includegraphics[width=.94\textwidth]{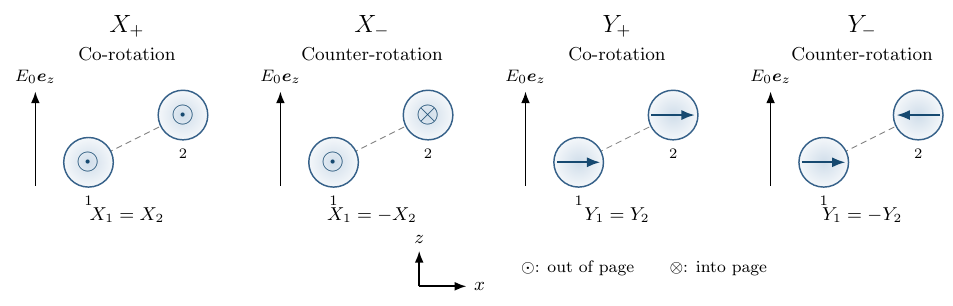}
\caption{The four leading rotation patterns near onset. The page is the
$x$--$z$ plane containing the line of centres and the imposed field.
Symbols and arrows indicate angular velocities: $\odot$ points out of
the page and $\otimes$ into the page; horizontal arrows point along
$\pm\boldsymbol e_x$. An $X$ mode has transverse dipoles along
$\pm\boldsymbol e_x$ and rotation axes normal to the page. A $Y$ mode has
transverse dipoles along $\pm\boldsymbol e_y$ and rotation axes in the page.
The $+$ and $-$ labels mean equal and opposite angular velocities,
respectively. Reversing both angular velocities gives the equivalent
state with the opposite sign of $a_k$ in Eq.~\eqref{eq:branchstates}.}
\label{fig:fourmodes}
\end{figure*}

The $+$ sign describes equal transverse dipoles on the two particles;
the $-$ sign describes opposite transverse dipoles.
After the initial transient, Eq.~\eqref{eq:fast} gives
$\boldsymbol\omega_i^{[1]}=\Mac^{-1}\boldsymbol a_i\times\ez$,
while Eq.~\eqref{eq:slaving} gives
$\boldsymbol\omega_i^{[2]}=\boldsymbol0$.
Using $\boldsymbol a_i=X_i\boldsymbol e_x+Y_i\boldsymbol e_y$,
the angular-velocity expansion therefore becomes
\begin{equation}
\begin{split}
\Om_i&=\frac{\eta}{\Mac}\boldsymbol a_i\times\ez+\order(\eta^3)\\
&=\frac{\eta}{\Mac}
(Y_i\boldsymbol e_x-X_i\boldsymbol e_y)+\order(\eta^3).
\label{eq:modeomega}
\end{split}
\end{equation}
Thus $X_+$ and $Y_+$ give co-rotation, while $X_-$ and $Y_-$
give counter-rotation, to leading order. An $X$ dipole component gives rotation about
$\boldsymbol e_y$, perpendicular to the pair--field plane. A $Y$
dipole component gives rotation about $\boldsymbol e_x$, within that
plane. Figure~\ref{fig:fourmodes} illustrates these four patterns.

\begin{figure*}
\centering
\includegraphics[width=.98\textwidth]{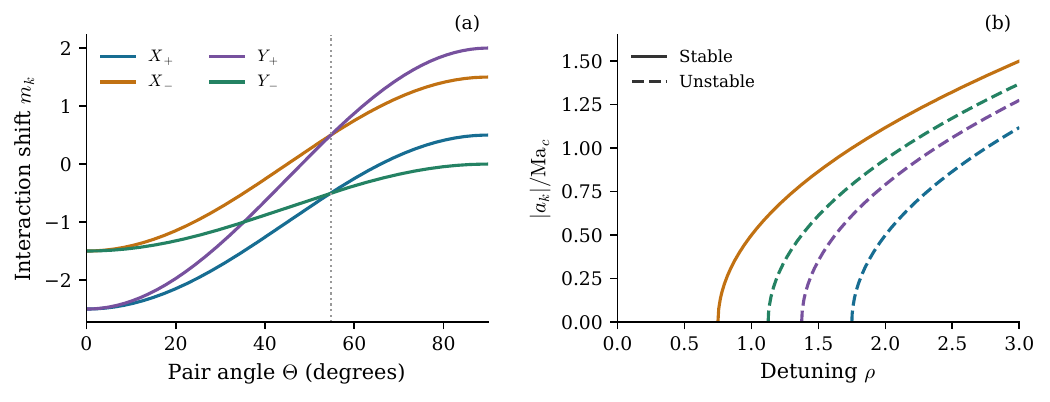}
\caption{Growth-rate shifts and steady amplitudes near onset for
$\epsc=-0.1097$ and $\sigc=-0.5$. (a) Interaction shifts $m_k$ for the
four modes. The vertical line marks
$\Theta_{\mathrm m}=\cos^{-1}(1/\sqrt3)$, where the largest shifts
$m_{X_-}$ and $m_{Y_+}$ coincide. (b) Scaled steady amplitude
$\lvert a_k\rvert/\Mac$ on either sphere at $\Theta=30^\circ$. Solid curves are
stable and dashed curves unstable to small perturbations of the four
amplitude components.
Colours in (b) follow the legend in (a).}
\label{fig:modes}
\end{figure*}

Setting $\Lambda_k=0$ gives the critical detuning
$\rho_c^{(k)}=-m_k$, so the largest shift identifies the lowest onset
field. Figure~\ref{fig:modes}(a) shows how the four interaction shifts
vary with pair angle. The largest shift is $m_{X_-}$ below
$\Theta_{\mathrm m}$ and $m_{Y_+}$ above it, so the first unstable mode
changes from counter-rotation to co-rotation at their crossing.
The electric interaction gives the same shift to all four growth rates;
the hydrodynamic interaction makes them different.
Since $\lambda_0=\rho\eta^2$, the corresponding critical field is
\begin{equation}
\left(\frac{E_{c,k}}{E_c^{(0)}}\right)^2
=1-\eta^2 m_k+\order(\eta^4).
\label{eq:modeonset}
\end{equation}
Equation~\eqref{eq:modeonset} is the leading near-onset asymptotic
prediction for the interaction-induced change in each modal critical
field. A positive $m_k$ lowers that field, whereas a negative $m_k$
raises it. Below, we compare these modal predictions with the stationary
thresholds of the unexpanded dipole--rotlet model and compare their
first-onset envelope with the boundary element results.

Since $T_2=\eta^2t$, a linear mode evolves as
\begin{equation}
e^{\Lambda_kT_2}=e^{\eta^2\Lambda_kt}
=e^{\eta^2\Lambda_k t_{\mathrm{phys}}/\tau_{\mathrm{MW}}}.
\label{eq:rateconversion}
\end{equation}
Its leading growth rate is therefore $\eta^2\Lambda_k$ per unit
dimensionless time, or $\eta^2\Lambda_k/\tau_{\mathrm{MW}}$ per second.

When $\Theta=0$, the line of centres and the field are parallel, so no
direction in the transverse plane is preferred. Accordingly,
$m_{X_\pm}=m_{Y_\pm}=4\sigc\mp1/2$. Because the corresponding $X$ and $Y$ growth rates coincide,
the rotation axes can take any common azimuth.

\subsection{Steady amplitudes and their stability}
\label{sec:amplitudestability}
For an $X$ mode, set $X_1=a_k$, $X_2=qa_k$, and $Y_1=Y_2=0$,
where $q=+1$ means equal leading transverse dipoles and $q=-1$
means opposite ones. Thus $q$ is the sign in the mode label $X_q$ or
$Y_q$; we also use this sign label in the onset and angular-speed
formulas below. For a $Y$ mode, set $Y_1=a_k$, $Y_2=qa_k$, and $X_1=X_2=0$. Substitution into
Eqs.~\eqref{eq:Xamp} and \eqref{eq:Yamp} gives the scalar equation
\begin{equation}
\frac{\dd a_k}{\dd T_2}=\Lambda_k a_k-\frac{a_k^3}{\Mac^2}.
\label{eq:branchscalar}
\end{equation}
Besides $a_k=0$, its steady solutions are
\begin{equation}
a_k=\pm \Mac\sqrt{\Lambda_k},\qquad \Lambda_k>0.
\label{eq:branchstates}
\end{equation}
Their magnitude grows continuously from zero as the field exceeds the
critical field for that mode. The two signs reverse both particles'
rotations. Here $a_k$ is the amplitude on one particle; the corresponding
sum or difference in Eq.~\eqref{eq:paircoords} has magnitude
$\sqrt2|a_k|$. To leading order the transverse dipole magnitude is
$\eta|a_k|$ and the angular speed is $\eta\sqrt{\Lambda_k}$.

Equation~\eqref{eq:branchscalar} describes perturbations that keep the chosen rotation pattern.
To establish stability we must also perturb the other dipole components.
For example, consider an $X_-$ state with
$(X_1,X_2)=(a_k,-a_k)$. In each row below, $\Delta X(T_2)$ or
$\Delta Y(T_2)$ denotes the infinitesimal amplitude of the indicated
perturbation. The rows are considered separately, each with its own
growth rate:
\begin{equation}
\begin{array}{c|c}
\text{perturbation}&\text{growth rate on }T_2\\ \hline
(\Delta X_1,\Delta X_2)=(\Delta X,-\Delta X)&-2\Lambda_{X_-}\\
(\Delta X_1,\Delta X_2)=(\Delta X,\Delta X)&\Lambda_{X_+}-3\Lambda_{X_-}\\
(\Delta Y_1,\Delta Y_2)=(\Delta Y,\Delta Y)&\Lambda_{Y_+}-\Lambda_{X_-}\\
(\Delta Y_1,\Delta Y_2)=(\Delta Y,-\Delta Y)&\Lambda_{Y_-}-\Lambda_{X_-}
\end{array}
\label{eq:branchspectrum}
\end{equation}
These rates follow by substituting the steady state plus the indicated
small perturbations into Eqs.~\eqref{eq:Xamp} and \eqref{eq:Yamp},
and retaining terms linear in $\Delta X$ and $\Delta Y$. The first perturbation changes
the common magnitude of the two opposite dipoles. The second adds equal
$x$ components, making their magnitudes unequal. The last two add
$y$ components and therefore change the directions of the rotation axes.
For example, the last pair obeys
$\dd(\Delta Y)/\dd T_2=(\Lambda_{Y_-}-\Lambda_{X_-})\Delta Y$.

A steady state is stable if all four growth rates are negative. The full set of growth rates for all four patterns is given in
Table~\hyperref[tab:spectra]{S1} in Sec.~\hyperref[sec:branchSM]{SIII} of the SM. It shows
that the mode with the unique largest $m_k$ gives a stable nonzero
state whenever $\rho+m_k>0$. At $\Theta=30^\circ$, $X_-$ has
the largest shift and therefore the lowest critical field.
Figure~\ref{fig:modes}(b) shows that its nonzero branch is stable
above this field (solid curve). The other single-mode branches
are unstable (dashed curves). Each branch
emerges continuously from zero at its own critical detuning and then
grows with the square root of the distance above that threshold. The
horizontal-axis intercepts therefore give the critical detuning of each
branch; solid and dashed lines denote stable and unstable nonzero
branches, respectively.

To determine which pattern first appears, substitute
Eqs.~\eqref{eq:mX} and \eqref{eq:mY} and cancel their common
electric contribution:
\begin{equation}
m_{X_-}-m_{X_+}=1,\qquad
m_{Y_+}-m_{X_-}=\tfrac12(1-3\cos^2\Theta).
\end{equation}
Also $m_{Y_-}\leq m_{X_-}$, with equality only at $\Theta=0$.
Thus counter-rotation of type $X_-$ first appears for
$0<\Theta<\Theta_{\mathrm m}$, and co-rotation of type $Y_+$ first
appears for $\Theta_{\mathrm m}<\Theta\leq\pi/2$, where
\begin{equation}
\Theta_{\mathrm m}=\cos^{-1}(1/\sqrt3)\simeq54.7^\circ.
\label{eq:magic}
\end{equation}
This angle also agrees with the small-amplitude limit of the
away-from-onset phase-selection boundary in Eq.~\eqref{eq:outerselection}:
$\cos^2\Theta_*=1/[3(1+\lambda_0)]$ tends to $1/3$ as
$\lambda_0\to0^+$. This agreement connects the rotating-state and
near-onset asymptotics in the matching limit $\delta\ll\lambda_0\ll1$.

At $\Theta=0$, rotating both axes by the same azimuthal
angle gives another equivalent counter-rotating state. At $\Theta=\Theta_{\mathrm m}$, the leading equations admit steady
solutions with both $X_-$ and $Y_+$ nonzero and the same total squared
amplitude. Higher-order terms are needed to determine selection within
this continuous family. The corresponding degeneracy is described in
Sec.~\hyperref[sec:branchSM]{SIII} of the SM.

\section{Separating electric and hydrodynamic interactions}
\label{sec:switched}
\subsection{Governing equations and non-rotating state}
To separate the effects of the two interaction mechanisms, introduce
$s_{\mathrm E},s_{\mathrm H}\in\{0,1\}$ and write
\begin{subequations}\label{eq:switchedmodel}
\begin{align}
\boldsymbol E_i&=\ez-s_{\mathrm E}\delta\PiR\cdot\Pp_j,\label{eq:switchedE}\\
\Om_i+s_{\mathrm H}\tfrac\delta2\PiR\cdot\Om_j
 &=\Ma^{-1}\Pp_i\times\boldsymbol E_i,\label{eq:switchedO}\\
\dot{\Pp}_i+s_{\mathrm E}\epsc\delta\PiR\cdot\dot{\Pp}_j
 &=\Om_i\times(\Pp_i-\epsc\boldsymbol E_i)
 -(\Pp_i-\sigc\boldsymbol E_i).\label{eq:switchedP}
\end{align}
\end{subequations}
The choices $(s_{\mathrm E},s_{\mathrm H})=(1,1),(1,0),(0,1),(0,0)$ give
both interactions, electric interaction only, hydrodynamic interaction only,
and isolated spheres, respectively. Switching the electric interaction off
removes the neighbor's contribution to the local field, its derivative, and
the induced change in the non-rotating state. The imposed dc field and the
electric torque responsible for Quincke rotation are retained in all four
cases.

The common non-rotating field and dipole are
\begin{equation}
\boldsymbol E_{\mathrm{nr}}
 =(\I+s_{\mathrm E}\sigc\delta\PiR)^{-1}\cdot\ez,
\qquad \Pp_{\mathrm{nr}}=\sigc\boldsymbol E_{\mathrm{nr}}.
\label{eq:switchedbase}
\end{equation}
Because $\Om_i=\boldsymbol0$ in this state, the rotlet flow vanishes.
The base field depends on $s_{\mathrm E}$ alone; with $s_{\mathrm E}=0$,
$\boldsymbol E_{\mathrm{nr}}=\ez$ and $\Pp_{\mathrm{nr}}=\sigc\ez$.

\subsection{Stationary onset in the unexpanded model}
\label{sec:stationaryonset}
We first obtain the stationary critical fields of the unexpanded model,
then compare their first-order expansion with the near-onset asymptotic
predictions. In the linear stability analysis of the non-rotating
state, write $\Pp_i=\Pp_{\mathrm{nr}}+\boldsymbol p_i$ and
$\Om_i=\boldsymbol\omega_i$, where the added quantities are
infinitesimal perturbations. Equation~\eqref{eq:switchedE} then gives
$\boldsymbol E_i=\boldsymbol E_{\mathrm{nr}}
-s_{\mathrm E}\delta\PiR\cdot\boldsymbol p_j$.
Thus the local-field perturbation is determined by the other dipole.
Adding and subtracting the two linearized particle equations gives
independent problems for equal and opposite perturbations,
$\boldsymbol p_2=q\boldsymbol p_1$ and
$\boldsymbol\omega_2=q\boldsymbol\omega_1$, with $q=+1$ or $-1$. We write $\boldsymbol\omega_q=\boldsymbol\omega_1$
for the angular-velocity perturbation in each sector.

At a stationary critical field the leading growth rate reaches zero.
The time derivatives in the linearized Eq.~\eqref{eq:switchedP}
then vanish, giving
$\boldsymbol p_i+s_{\mathrm E}\sigc\delta\PiR\cdot\boldsymbol p_j
=\Mac\boldsymbol E_{\mathrm{nr}}\times\boldsymbol\omega_i$.
Substituting this relation into the linearized
Eq.~\eqref{eq:switchedO}, and using
$\Ma^{-1}=(E_0/E_c^{(0)})^2/\Mac$ from Eq.~\eqref{eq:lambda},
gives the neutral angular-velocity equation
\begin{equation}
\begin{split}
\left(\I+\frac{q s_{\mathrm H}\delta}{2}\PiR\right)
 \cdot\boldsymbol\omega_q
={}&\left(\frac{E_0}{E_c^{(0)}}\right)^2
 \bigl[|\boldsymbol E_{\mathrm{nr}}|^2\I\\
&-\boldsymbol E_{\mathrm{nr}}\boldsymbol E_{\mathrm{nr}}\bigr]
 \cdot\boldsymbol\omega_q.
\end{split}
\label{eq:neutralmatrix}
\end{equation}
For each sign of $q$, this equation gives two positive critical fields,
corresponding to two independent rotation-axis directions. For sufficiently
small $\delta$, the linearized time-dependent equations have real growth
rates, so onset occurs through a growth rate crossing zero. The linearized equations and the proof that all growth rates are real
are given in Secs.~\hyperref[sec:linearSM]{SIV\,A.1}--\hyperref[sec:realitySM]{A.2}
of the SM.

\begin{figure*}
\includegraphics[width=.98\textwidth]{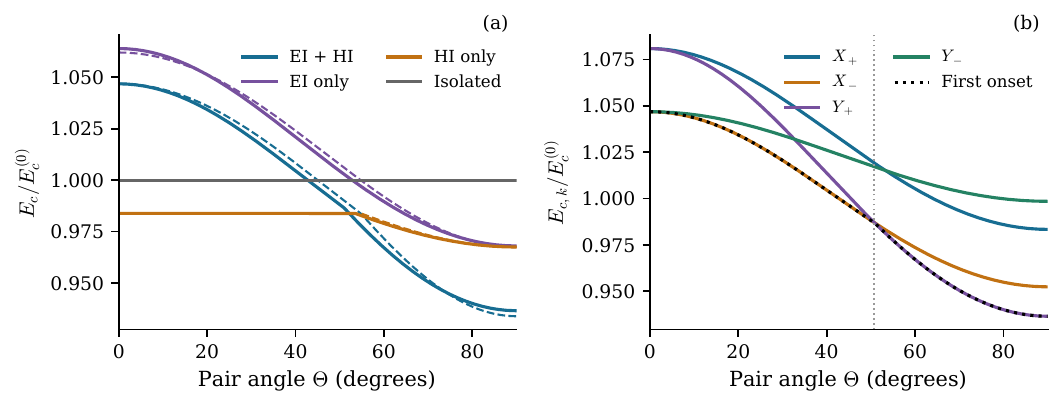}
\caption{Stationary onset fields of the dipole--rotlet model at
$R=2.5$, $\epsc=-0.1097$, and $\sigc=-0.5$. (a) First onset with both
interactions (EI + HI), electric interaction only (EI), hydrodynamic
interaction only (HI), and isolated spheres. Solid curves are the
unexpanded-model thresholds and dashed curves are the leading near-onset
asymptotic prediction~\eqref{eq:onsetenvelope}. (b) Four modal thresholds with both interactions.
The dotted black curve gives the lowest of the four thresholds at each
angle. The vertical dotted line marks the
unexpanded-model $X_-$--$Y_+$ crossover at
$\Theta_{\mathrm{cross}}=50.8^\circ$.
}
\label{fig:switches}
\end{figure*}

Because $\boldsymbol E_{\mathrm{nr}}$ and $\widehat{\boldsymbol R}$
lie in the $x$--$z$ plane, the matrix equation separates into rotation
normal to that plane and rotation within it.
Here $X_q$ denotes $X_+$ for $q=+1$ and $X_-$ for $q=-1$, and
likewise for $Y_q$. These finite-separation modes connect to the four
near-onset modes of Sec.~\ref{sec:modes}. The $X_q$ angular velocity is
normal to the pair--field plane, whereas
the $Y_q$ angular velocity lies within it. The calculation in
Sec.~\hyperref[sec:fourfieldsSM]{SIV\,A.3} of the SM evaluates Eq.~\eqref{eq:neutralmatrix} and
gives
\begin{equation}
\left(\frac{E_{c,X_q}}{E_c^{(0)}}\right)^2
 =\frac{1+q s_{\mathrm H}\delta/2}{|\boldsymbol E_{\mathrm{nr}}|^2},
\label{eq:exactX}
\end{equation}
\begin{equation}
\left(\frac{E_{c,Y_q}}{E_c^{(0)}}\right)^2
 =\frac{(1+q s_{\mathrm H}\delta/2)(1-q s_{\mathrm H}\delta)}
 {(1+q s_{\mathrm H}\delta/2)|\boldsymbol E_{\mathrm{nr}}|^2
 -(3q s_{\mathrm H}\delta/2)(\widehat{\boldsymbol R}\cdot\boldsymbol E_{\mathrm{nr}})^2}.
\label{eq:exactY}
\end{equation}
These expressions are exact within the unexpanded dipole--rotlet model.
The first onset occurs at the smallest of the four positive modal fields.
The local-field derivative term in Eq.~\eqref{eq:switchedP} affects
nonzero growth rates but vanishes in the stationary neutral condition.
Normalization by $E_c^{(0)}$ removes $\epsc$ from
Eqs.~\eqref{eq:exactX} and \eqref{eq:exactY}, so these normalized stationary
thresholds depend on the conductivity contrast through
$\boldsymbol E_{\mathrm{nr}}$.
Carrying the switches through the multiple-scale calculation of
Sec.~\ref{sec:multiple} multiplies the electric and hydrodynamic
contributions by $s_{\mathrm E}$ and $s_{\mathrm H}$, respectively.
The resulting leading near-onset asymptotic growth-rate shifts are
\begin{align}
m_{X_\pm}&=-2s_{\mathrm E}\sigc(1-3\cos^2\Theta)
 \mp\frac{s_{\mathrm H}}2,\label{eq:switchedmX}\\
m_{Y_\pm}&=-2s_{\mathrm E}\sigc(1-3\cos^2\Theta)
 \pm\frac{s_{\mathrm H}}2(2-3\cos^2\Theta).
\label{eq:switchedmY}
\end{align}
For comparison with the unexpanded model, recall $\delta=\eta^2$.
Both calculations start from the same dipole--rotlet equations.
Expanding Eqs.~\eqref{eq:exactX} and \eqref{eq:exactY} to first order
in $\delta$ gives the same coefficients $m_k$ as the near-onset
multiple-scale calculation, for every switch setting. This agreement
provides a nontrivial consistency check between the stationary linear
stability calculation and the detuned multiple-scale derivation.
Equation~\eqref{eq:modeonset} gives each modal squared critical field.
The lowest of these fields corresponds to the largest $m_k$;
selecting it and taking the square root gives

\begin{equation}
\frac{E_c}{E_c^{(0)}}
 =\sqrt{1-\delta\max_k m_k}+\order(\delta^2).
\label{eq:onsetenvelope}
\end{equation}
Equation~\eqref{eq:onsetenvelope} therefore contains the same first-order
asymptotic information as Eq.~\eqref{eq:modeonset}, with the switches
included and the first mode selected. Figure~\ref{fig:switches}(a) separates the two mechanisms: electric
interaction alone raises onset near a field-aligned pair and lowers it
near a perpendicular pair for this material, while hydrodynamic coupling
splits the competing rotation modes. Their combined effect depends on pair angle. Panel (b) resolves the
four modal fields and identifies which one sets the first-onset envelope.
The solid and dashed curves remain close because the leading asymptotic
shifts are common to both calculations; their differences arise from
higher powers of $\delta$ retained in the unexpanded model.

\subsection{Interaction mechanisms and finite-separation crossover}
\label{sec:crossover}
We now use the unexpanded thresholds
\eqref{eq:exactX}--\eqref{eq:exactY} to distinguish the two interaction
mechanisms and determine how finite separation changes the crossover
between the first unstable modes.
With electric interaction only, $s_{\mathrm H}=0$ and all four stationary
thresholds coincide at $E_c/E_c^{(0)}=|\boldsymbol E_{\mathrm{nr}}|^{-1}$.
Electric interaction changes the local field magnitude and its direction,
but does not split these stationary thresholds within the unexpanded model.
With hydrodynamic interaction only, $\boldsymbol E_{\mathrm{nr}}=\ez$,
so the $X_q$ squared thresholds reduce to $1+q\delta/2$; the $Y_q$
thresholds follow Eq.~\eqref{eq:exactY} with
$|\boldsymbol E_{\mathrm{nr}}|=1$ and
$\widehat{\boldsymbol R}\cdot\boldsymbol E_{\mathrm{nr}}=\cos\Theta$.
The $X_-$ onset is therefore independent of angle until the $Y_+$ mode
sets a lower threshold at larger pair angles.

When the hydrodynamic interaction is present, the first instability is
controlled by $X_-$ or $Y_+$. Neither $X_+$ nor $Y_-$ becomes
unstable at a lower field than $X_-$. Determining the first onset
therefore requires only a comparison of $X_-$ and $Y_+$.
Let $\Theta_{\mathrm{cross}}(R)$ denote the pair angle where the
$X_-$ and $Y_+$ onset fields coincide. Equating these two thresholds gives
\begin{equation}
\frac{(\widehat{\boldsymbol R}\cdot\boldsymbol E_{\mathrm{nr}})^2}
 {|\boldsymbol E_{\mathrm{nr}}|^2}
 =\frac{1+\delta/2}{3(1-\delta/2)}.
\label{eq:crosslocal}
\end{equation}
In the leading near-onset asymptotic limit, this reduces to $\cos^2\Theta=1/3$. At finite separation,
the higher powers of $\delta$ retained by the unexpanded model and
the electric tilt of the base field shift the angle at which the two
critical fields coincide. With hydrodynamic interaction only,
\begin{equation}
\cos^2\Theta_{\mathrm{cross}}^{\mathrm H}(R)
 =\frac{1+1/(2R^3)}{3[1-1/(2R^3)]}.
\label{eq:HIcross}
\end{equation}
The left-hand side of Eq.~\eqref{eq:crosslocal} is the squared cosine
of the angle between the line of centres and the local non-rotating
electric field. With hydrodynamic interaction alone, this field is the
imposed field, so the left-hand side is $\cos^2\Theta$. Electric interaction
tilts the local field. The same criterion on its orientation relative to
the pair axis is therefore reached at a different value of $\Theta$,
which is measured from the imposed field.
For $\sigc=-0.5$ and $R=2.5$, this
gives $\Theta_{\mathrm{cross}}=50.8^\circ$, compared with
$\Theta_{\mathrm{cross}}^{\mathrm H}=53.4^\circ$ and the near-onset asymptotic value
$\Theta_{\mathrm m}=54.7^\circ$. At $R=6$, the corresponding values
are $54.45^\circ$ and $54.64^\circ$, approaching the near-onset asymptotic angle.
The kink in Fig.~\ref{fig:switches}(a,b) results from taking the lower
of the smooth $X_-$ and $Y_+$ threshold curves. The envelope is continuous,
but their unequal slopes produce a change in its angular derivative at
the crossing. Both the unexpanded model and the leading asymptotic theory
predict this switch of the first unstable mode, generally at different
angles. The explicit crossover calculation is given in
Sec.~\hyperref[sec:crossoverSM]{SIV\,A.4} of the SM.

\section{Steady rotation away from onset}
\label{sec:finiteamp}
The near-onset amplitude equations apply when
$\lambda_0=\order(\delta)$. We now return to the regular expansion
of Sec.~\ref{sec:regular}, which gives steady angular-speed predictions
away from onset. This calculation holds at a fixed supercritical field
with $\lambda_0>0$ and requires $\delta/\lambda_0\ll1$, so that the
interaction correction remains small relative to the isolated rotation.
At $E_0/E_c^{(0)}=2$, the phase-selection boundary in
Eq.~\eqref{eq:outerselection} is $\Theta_*\simeq73.2^\circ$.
Counter-rotation is stable below this angle and the co-rotating branch
of Eq.~\eqref{eq:outercorotation} is stable above it. We therefore evaluate the
counter-rotating state $(\phi_1,\phi_2)=(0,\pi)$ at
$\Theta=0,\pi/4$, and the co-rotating state with
$\phi_1=\phi_2=\pi/2$ at $\Theta=\pi/2$.
Their transverse dipoles correspond to the $X_-$ and $Y_+$ patterns
of Sec.~\ref{sec:modes}, respectively.

For the theory and simulation comparisons, define the mean angular speed
\begin{equation}
\overline\Omega=\frac{|\Om_1|+|\Om_2|}{2}.
\label{eq:meanspeed}
\end{equation}
We impose Eq.~\eqref{eq:phaseforcing}, substitute
the steady dipole corrections \eqref{eq:steadyz} and
\eqref{eq:steadyrho} into Eq.~\eqref{eq:Ofirst}, and obtain
the first-order regular asymptotic predictions
\begin{equation}
\begin{split}
\overline\Omega_{X_-}={}&\sqrt{\lambda_0}
 -\delta\frac{(\Pperp^{(0)})^2+\Ma^2}{\Pperp^{(0)}\Ma}\\
&\times\left[(\epsc-\Ma)(1-3\cos^2\Theta)-\tfrac14\right]
 +\order(\delta^2),
\end{split}
\label{eq:steadyXm}
\end{equation}
\begin{equation}
\begin{split}
\overline\Omega_{Y_+}\big|_{\Theta=\pi/2}
={}&\sqrt{\lambda_0}-\delta\frac{(\Pperp^{(0)})^2+\Ma^2}{\Pperp^{(0)}\Ma}\\
&\times\left(\epsc-\Ma-\tfrac12\right)+\order(\delta^2).
\end{split}
\label{eq:steadyYp}
\end{equation}
In each bracket, the term containing $\epsc-\Ma$ comes from the
electric dipole field of the other sphere; the constant $-1/4$ or
$-1/2$ comes from its rotlet flow. Their signs determine whether each
interaction speeds up or slows down the chosen rotating state.
The component derivation and overlap with the near-onset amplitudes
are given in Sec.~\hyperref[sec:steadyfinite]{SIV\,B} of the SM. Section~\ref{sec:numerics}
compares these regular asymptotic mean speeds with the unexpanded
model and BEM results.

\section{Boundary element formulation}
\label{sec:bem}
\begin{figure}[!t]
\centering
\includegraphics[width=\columnwidth]{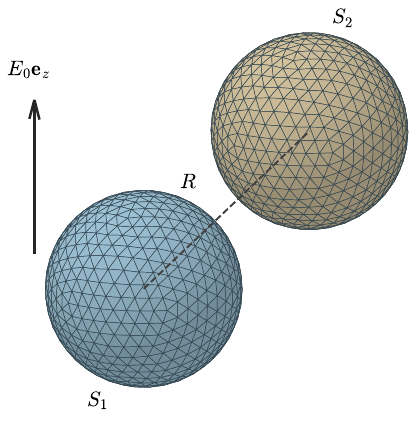}
\caption{Discretized sphere pair at $R=2.5$ and $\Theta=\pi/4$.
Each surface contains $642$ nodes and $1280$ flat three-node triangles,
generated by three subdivisions of an icosahedron. Surface densities are
interpolated with continuous piecewise-linear (P1) basis functions.
The arrow indicates the imposed field; the dashed line joins the centres.}
\label{fig:bemmesh}
\end{figure}
The boundary element method~\cite{pozrikidis1992boundary,pozrikidis2002practical}
is well suited to simulating the electrohydrodynamics of particles.
It solves Laplace's equation for the electric
potential~\cite{das2017nonlinear,DasSaintillan2017} and the Stokes
equations for the fluid flow~\cite{das2018computing,das2019transition}
through boundary integral equations, requiring only the particle
surfaces to be discretized. Here, we extend the validated
single-particle solver~\cite{DasLauga2019} to two interacting rigid
spheres. The simulations evolve the surface charge and resolve the
coupled electric and hydrodynamic interactions, providing a numerical
benchmark for the dipole--rotlet model and asymptotic theory.
For the Stokes problem, we use the completed, deflated double-layer
formulation of Pozrikidis~\cite{Pozrikidis2006}.
Figure~\ref{fig:bemmesh} shows the flat triangular mesh used to
discretize each sphere.

To extend the validated single-particle solver to two spheres with
minimal changes, we retain its electroviscous time scale
$\tau_E=\mu/(\epsilon_1E_0^2)$. For dimensional time $t_{\mathrm{phys}}$
and angular velocity $\Om_{\mathrm{phys}}$, the solver uses
$t_B=t_{\mathrm{phys}}/\tau_E$ and $\Om_B=\tau_E\Om_{\mathrm{phys}}$.
We express the results and the equations below in the Maxwell--Wagner
units of Sec.~\ref{sec:setup}. Since $\Ma=2\tau_E/\tau_{\mathrm{MW}}$,
the conversion is
\begin{equation}
t=\frac{\Ma}{2}t_B,\qquad \Om=\frac{2}{\Ma}\Om_B.
\label{eq:bemtimeconversion}
\end{equation}

Let $S=S_1\cup S_2$, with the unit normal pointing from each solid into
the fluid, and let $\boldsymbol s_i(\boldsymbol x)=\boldsymbol x-\boldsymbol x_{c,i}$
denote the displacement from the centre of sphere $i$. We scale length
by $a$, electric field by $E_0$ and free surface charge by $\epsilon_1E_0$.
The physical material ratios are
\begin{equation}
Q=\frac{\epsilon_2}{\epsilon_1},\qquad
\kappa=\frac{\sigma_2}{\sigma_1},\qquad
\chi=\frac{Q+2}{\kappa+2}.
\label{eq:bemratios}
\end{equation}
Superscripts $+$ and $-$ denote electric fields evaluated at the
interface on the fluid and solid sides, respectively. Continuity of potential and Gauss's law give
$\boldsymbol E_t^+=\boldsymbol E_t^-$ and $q_s=E_n^+-QE_n^-$.
For prescribed surface charge, the electrical unknown is the jump in
normal electric field,
\begin{equation}
\Delta E_{n,i}=E_{n,i}^+-E_{n,i}^-.
\label{eq:normaljump}
\end{equation}
It is also the density of the single-layer potential. With
$G(\boldsymbol x_0,\boldsymbol x)=[4\pi|\boldsymbol x_0-\boldsymbol x|]^{-1}$,
the potential is
\begin{equation}
\Phi(\boldsymbol x_0)=-\ez\cdot\boldsymbol x_0+
\sum_{j=1}^2\int_{S_j}G(\boldsymbol x_0,\boldsymbol x)
\Delta E_{n,j}(\boldsymbol x)\,\dd S.
\label{eq:bemphi}
\end{equation}
For a target $\boldsymbol x_0\in S_i$, the regularized equation for this
jump is
\begin{equation}
\begin{split}
&\int_{S_i}[\Delta E_{n,i}(\boldsymbol x)-\Delta E_{n,i}(\boldsymbol x_0)]
\boldsymbol n_i(\boldsymbol x_0)\cdot\nabla_0G\,\dd S\\
&+\sum_{j\ne i}\int_{S_j}\Delta E_{n,j}(\boldsymbol x)
\boldsymbol n_i(\boldsymbol x_0)\cdot\nabla_0G\,\dd S\\
&+\Delta E_{n,i}(\boldsymbol x_0)
\left[\frac{Q}{Q-1}-L_i(\boldsymbol x_0)\right]\\
&\qquad=\ez\cdot\boldsymbol n_i(\boldsymbol x_0)
+\frac{q_{s,i}(\boldsymbol x_0)}{Q-1}.
\end{split}
\label{eq:bemelectric}
\end{equation}
Subtracting the target value of the density regularizes the self-surface
integral. The integral over the other sphere has no coincident-point
singularity. The geometric correction from the self-surface subtraction
of Sellier~\cite{Sellier2006} and Das and Saintillan~\cite{DasSaintillan2017}
takes the following form on sphere $i$,
\begin{equation}
\begin{split}
L_i(\boldsymbol x_0)={}&
\boldsymbol n_i(\boldsymbol x_0)\cdot\int_{S_i}
\Bigl\{2G\,\boldsymbol n_i(\boldsymbol x)\\
&+[\boldsymbol n_i(\boldsymbol x)-\boldsymbol n_i(\boldsymbol x_0)]
[\boldsymbol n_i(\boldsymbol x)\cdot\nabla_{\boldsymbol x}G]\Bigr\}\,\dd S .
\end{split}
\label{eq:bemselfgeometry}
\end{equation}
Here $\nabla_{\boldsymbol x}G$ differentiates with respect to the
integration point $\boldsymbol x$. The factor $2$ is the surface
divergence of the outward unit normal on a sphere of unit radius.

After solving Eq.~\eqref{eq:bemelectric}, the normal fields follow from
\begin{equation}
E_{n,i}^+=\frac{q_{s,i}-Q\Delta E_{n,i}}{1-Q},\qquad
E_{n,i}^-=\frac{q_{s,i}-\Delta E_{n,i}}{1-Q}.
\label{eq:normalfields}
\end{equation}
The surface potential is then evaluated using Eq.~\eqref{eq:bemphi},
and its surface gradient gives
\begin{equation}
\boldsymbol E_{t,i}=-\boldsymbol\nabla_s\Phi.
\label{eq:tangentialfield}
\end{equation}
This avoids direct evaluation of the strongly singular tangential-field
integral. The reconstructed fields determine the dimensionless
Maxwell-stress jump and electric torque:
\begin{equation}
\begin{split}
\boldsymbol f_{E,i}={}&\bigl[\boldsymbol E_i^+\boldsymbol E_i^+
-\tfrac12|\boldsymbol E_i^+|^2\I\\
&-Q\bigl(\boldsymbol E_i^-\boldsymbol E_i^-
-\tfrac12|\boldsymbol E_i^-|^2\I\bigr)\bigr]\cdot\boldsymbol n_i,
\end{split}
\label{eq:bemtraction}
\end{equation}
\begin{equation}
\boldsymbol T_{E,i}=\int_{S_i}\boldsymbol s_i(\boldsymbol x)
\times\boldsymbol f_{E,i}\,\dd S.
\label{eq:bemtorque}
\end{equation}
The steady non-rotating charge distribution gives zero net electric
torque on each sphere. Numerical integration on the mesh leaves a small
residual $\boldsymbol T_{E,i}^{\mathrm{nr}}$. We evaluate this residual
once on the unperturbed state and subtract it from each subsequent
torque so that a perturbation of the charge distribution drives the
measured rotation.

The exterior fluid obeys the incompressible Stokes equations and is at
rest at infinity. To isolate torque-driven rotational coupling, we
prescribe the electric torque on each sphere and set both force inputs
to zero in the Stokes mobility problem. Let $\boldsymbol\psi_j$ be the
Stokes double-layer density on $S_j$. The completed double-layer
representation~\cite{PowerMiranda1987,Pozrikidis2006} gives the exterior
velocity at $\boldsymbol x_0$ as
\begin{subequations}\label{eq:bemdlp}
\begin{equation}
\boldsymbol u(\boldsymbol x_0)=
\sum_{j=1}^2\int_{S_j}\mathsf K_j\cdot\boldsymbol\psi_j(\boldsymbol x)\,\dd S
+\boldsymbol v(\boldsymbol x_0).
\label{eq:bemstokesfield}
\end{equation}
At the integration point $\boldsymbol x\in S_j$, let
$\boldsymbol r=\boldsymbol x-\boldsymbol x_0$. Contracting the stresslet
with the outward unit normal $\boldsymbol n_j(\boldsymbol x)$ gives
\begin{equation}
\mathsf K_j(\boldsymbol x,\boldsymbol x_0)=
-6\,\frac{\boldsymbol r\cdot\boldsymbol n_j(\boldsymbol x)}
{|\boldsymbol r|^5}\,\boldsymbol r\boldsymbol r .
\label{eq:bemstresslet}
\end{equation}
The point torques at the sphere centres give the completion velocity
\begin{equation}
\boldsymbol v(\boldsymbol x_0)=
\frac{1}{4\pi\Ma}\sum_{j=1}^2
\frac{(\boldsymbol T_{E,j}-\boldsymbol T_{E,j}^{\mathrm{nr}})
\times\boldsymbol s_j(\boldsymbol x_0)}
{|\boldsymbol s_j(\boldsymbol x_0)|^3}.
\label{eq:bemrotlets}
\end{equation}
\end{subequations}
The prefactor uses the Maxwell--Wagner time scale and the Mason number
in Eq.~\eqref{eq:Ma}, which is twice that used in Ref.~\cite{DasLauga2019}.

At a target $\boldsymbol x_0\in S_i$, the no-slip condition is
\begin{subequations}\label{eq:bemmobility}
\begin{equation}
\boldsymbol u(\boldsymbol x_0)=\boldsymbol U_i+
\Om_i\times\boldsymbol s_i(\boldsymbol x_0).
\label{eq:bemrigid}
\end{equation}
Let $A_i$ denote the area of $S_i$. Applying the self-surface subtraction
to the deflated boundary integral equation of Pozrikidis~\cite{Pozrikidis2006}
gives
\begin{equation}
\begin{aligned}
 &-\int_{S_i}\mathsf K_i\cdot
 [\boldsymbol\psi_i(\boldsymbol x)-\boldsymbol\psi_i(\boldsymbol x_0)]\,\dd S\\
 &-\sum_{j\ne i}\int_{S_j}\mathsf K_j\cdot\boldsymbol\psi_j(\boldsymbol x)\,\dd S\\
 &+\frac{4\pi}{A_i}\,\boldsymbol n_i(\boldsymbol x_0)
 \int_{S_i}\boldsymbol\psi_i\cdot\boldsymbol n_i\,\dd S\\
 &+\boldsymbol U_i+\Om_i\times\boldsymbol s_i(\boldsymbol x_0)
 =\boldsymbol v(\boldsymbol x_0).
\end{aligned}
\label{eq:bemdeflated}
\end{equation}
The local term $-4\pi\boldsymbol\psi_i(\boldsymbol x_0)$ in the subtraction
identity cancels the exterior jump, leaving the density difference in the
first integral.
The second integral gives the contribution from the other sphere and
is regular. The term proportional to $\boldsymbol n_i(\boldsymbol x_0)$
deflates the normal mode on $S_i$.

The translational and angular velocities in Eq.~\eqref{eq:bemdeflated}
follow from the surface moments of the same density~\cite{Pozrikidis2006},
\begin{align}
\boldsymbol U_i&=-\frac{4\pi}{A_i}\int_{S_i}\boldsymbol\psi_i\,\dd S,
\label{eq:bemstokessurfaceaverage}\\
\mathsf I_i\cdot\Om_i&=-4\pi\int_{S_i}\boldsymbol s_i(\boldsymbol x)\times
\boldsymbol\psi_i\,\dd S,
\label{eq:bemstokesangularmoment}\\
\mathsf I_i&=\int_{S_i}
\bigl(|\boldsymbol s_i(\boldsymbol x)|^2\I-
\boldsymbol s_i(\boldsymbol x)\boldsymbol s_i(\boldsymbol x)\bigr)\,\dd S.
\label{eq:bemstokessurfaceinertia}
\end{align}
\end{subequations}
Here $\mathsf I_i$ is the surface moment of inertia tensor.
We solve Eq.~\eqref{eq:bemdeflated} for both densities together.
For the fixed pair we discard the computed translations and use the
angular velocities to advance the charge and material surface nodes.
\begin{align}
\frac{Dq_{s,i}}{Dt}
&=\chi(\kappa E_{n,i}^- -E_{n,i}^+),\label{eq:bemcharge}\\
\frac{\dd\boldsymbol x}{\dd t}
&=\Om_i\times\boldsymbol s_i(\boldsymbol x),
\quad\boldsymbol x\in S_i.
\label{eq:bemnodes}
\end{align}
The derivative follows material surface nodes. Rigid rotation preserves
area, so their motion incorporates charge convection.

The surfaces use flat P1 triangles, with continuous piecewise-linear
densities and $642$ nodes and $1280$ triangles on each sphere for the main
results. Singular element integrals use Duffy-transformed
quadrature~\cite{Duffy1982}, and the linear systems are solved with
GMRES~\cite{SaadSchultz1986}. Tangential fields are
obtained by local surface differentiation of the potential. Charge and
orientation are advanced together by explicit midpoint stepping with
$\Delta t_B=0.1$.
The dipole coefficient corresponding to the unexpanded-model convention
is recovered from the single-layer density as
\begin{equation}
\Pp_i=\frac{1}{4\pi}\int_{S_i}\Delta E_{n,i}(\boldsymbol x)
\boldsymbol s_i(\boldsymbol x)\,\dd S.
\label{eq:bemdipole}
\end{equation}
For the counter-rotating state at $R=8$, $\Theta=0$, and
$E_0/E_c^{(0)}=2$, with the material parameters in
Eq.~\eqref{eq:bemfarfieldmaterial}, increasing the mesh from $642$ to
$2562$ nodes per sphere changed the steady mean angular speed by
$0.035\%$ and the interaction-induced speed correction by approximately
$2\%$. Halving $\Delta t_B$ from $0.1$ to $0.05$ changed both quantities
by less than $0.001\%$. The interaction correction uses an isolated-sphere
baseline computed with matching numerical settings.
Separately, tightening both dipole-ODE integration tolerances by a
factor of $100$ changed the sampled dimensionless dipole components by
less than $7\times10^{-11}$ in tests at $R=2.5,4$ and
$\Theta=0,\pi/4,\pi/2$, with the material in Eq.~\eqref{eq:bemmaterial}.

\section{Numerical results}
\label{sec:numerics}
\subsection{Onset envelope from perturbation growth and decay}
The onset calculations use the material ratios
$Q\simeq1.4216$ and $\kappa=10^{-6}$, giving
\begin{equation}
\epsc\simeq0.1232,\qquad \sigc\simeq-0.5.
\label{eq:bemmaterial}
\end{equation}
This is a nearly insulating particle with higher permittivity than the
suspending fluid. For each comparison, the BEM simulations, unexpanded dipole--rotlet
model and asymptotic predictions use matching material parameters. The
angular-speed comparisons below use the second material in
Eq.~\eqref{eq:bemfarfieldmaterial}. The small finite conductivity
approximates an insulating sphere in both sets of simulations.

We locate onset by following small perturbations of the interacting
non-rotating charge distribution. For each geometry, a lower test field
is placed $3\%$ below the smaller of the unexpanded-model and
near-onset asymptotic first-onset fields, and an upper test field
$3\%$ above the larger of the two. All four rotation patterns
decay at the lower field, while at least one grows at the upper field.
The two fields therefore define an interval containing the onset of
rotation. These tests provide an initial comparison with the predicted
critical fields. Refining the interval by time integration becomes costly
near onset, because the leading growth or decay rate approaches zero and
an appreciable change takes progressively longer to observe. A
steady-state BEM solver with pseudo-arclength continuation and
bifurcation detection could locate onset directly and follow both
stable and unstable rotating branches. Each continuation step would
require a nonlinear solve coupling surface charge and rigid-body
rotation, making an efficient implementation a substantial computational
undertaking.

\begin{figure*}
\includegraphics[width=.98\textwidth]{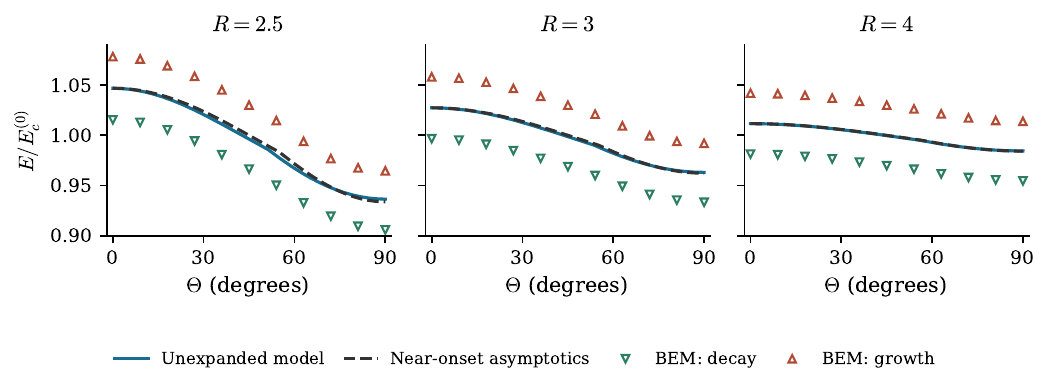}
\caption{Finite-resolution onset range with boundary element simulations for the material
ratios in Eq.~\eqref{eq:bemmaterial}. Solid curves are the stationary
thresholds of the unexpanded model; dashed curves are the leading near-onset
asymptotic prediction. Downward and upward open triangles denote BEM fields at which
all four tested rotation patterns decay or at least one grows, respectively. BEM simulations are shown at $9^\circ$ angular intervals.}
\label{fig:bemonset}
\end{figure*}

Figure~\ref{fig:bemonset} compares these intervals with the near-onset
asymptotic and unexpanded-model onset curves at $R=2.5,3,4$. The interaction lowers
or raises the onset field according to pair orientation. Its effect is
most visible for the closer pairs and weakens as the separation grows.
At the numerical resolution used, intervals entirely above
$E_0/E_c^{(0)}=1$ show that interactions raise onset, while intervals
entirely below unity show that they lower it. The former occur for
$R=2.5$ near $\Theta=0$; the latter occur for $R=2.5$ and $R=3$
near $\Theta=\pi/2$. At $R=4$, the intervals contain unity, so the
tests cannot determine the sign of the smaller shift.

The near-onset asymptotic curve and the stationary onset curve of the
unexpanded dipole--rotlet model approach one another as the separation increases, consistent with weak interactions.

\subsection{Growth and relaxation to steady rotation}
To compare directly with the material parameters used in
Ref.~\cite{DasSaintillan2013}, the angular-speed calculations use
\begin{equation}
\epsc=-0.1097,\qquad \sigc\simeq-0.5.
\label{eq:bemfarfieldmaterial}
\end{equation}
The field is $E_0/E_c^{(0)}=2$, so $\lambda_0=3$ and the
isolated-sphere angular speed is $\sqrt3$. We consider
$R=4,6,8$ at $\Theta=0,\pi/4,\pi/2$. The initial charge
perturbations select the stable $X_-$ and $Y_+$ patterns identified
in Sec.~\ref{sec:finiteamp}: counter-rotation for the first two
angles and co-rotation when the line of centres is perpendicular to the field.

We compare the mean angular speed defined in Eq.~\eqref{eq:meanspeed}.
The unexpanded equations are initialized about their interacting non-rotating state using
the initial dipole perturbation measured from BEM. Time and
angular velocity are expressed in the same Maxwell--Wagner
units for all three descriptions.

\begin{figure*}
\centering
\includegraphics[width=.98\textwidth]{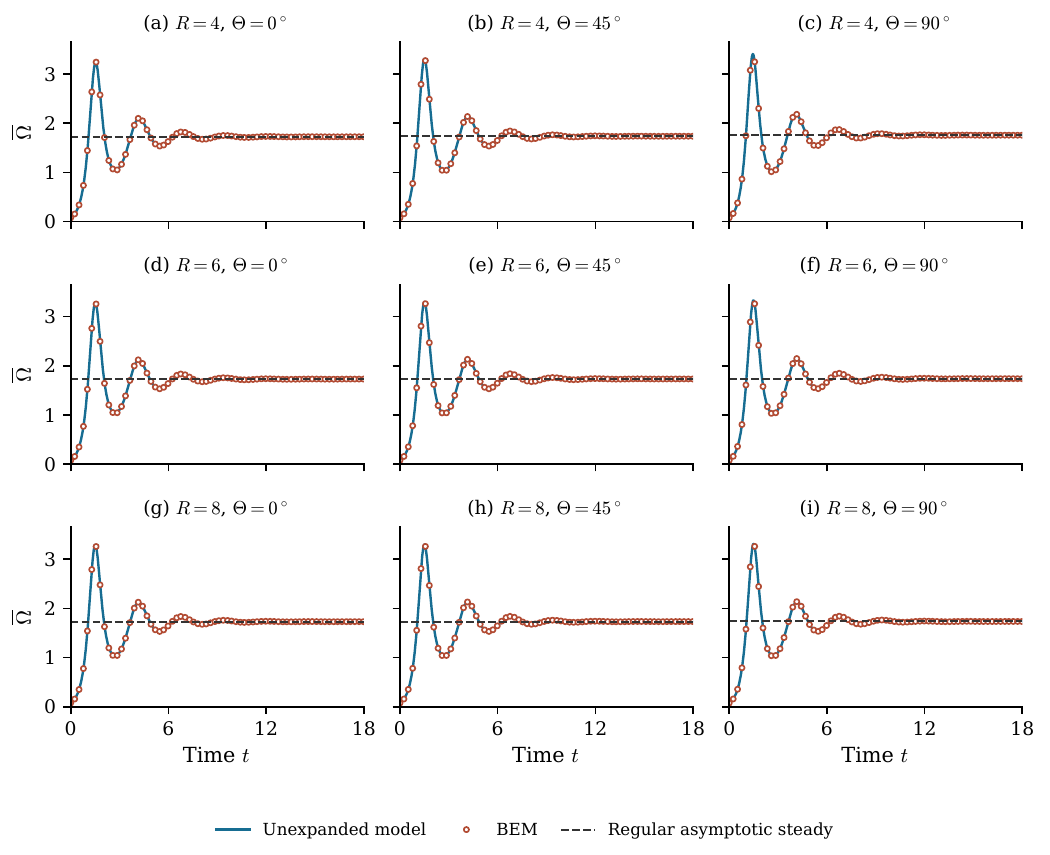}
\caption{Temporal evolution of the mean angular speed at
$E_0/E_c^{(0)}=2$ for the material in
Eq.~\eqref{eq:bemfarfieldmaterial}. Rows correspond to
$R=4,6,8$, and columns to $\Theta=0,\pi/4,\pi/2$.
The first two columns show counter-rotation of type $X_-$;
the last shows co-rotation of type $Y_+$.
Open circles are BEM simulations, solid curves are independent
integrations of the unexpanded dipole--rotlet model, and
horizontal dashed lines are the first-order regular asymptotic predictions.
The observable is defined in Eq.~\eqref{eq:meanspeed}.
Time is scaled by $\tau_{\mathrm{MW}}$ and angular velocity by
$\tau_{\mathrm{MW}}^{-1}$. BEM symbols are subsampled for clarity.}
\label{fig:bemomega}
\end{figure*}

Figure~\ref{fig:bemomega} shows the evolution of the mean angular speed
from the perturbed non-rotating state. BEM and the unexpanded
model exhibit the same rapid growth, overshoot, and damped
approach to steady rotation. The horizontal dashed lines are
the first-order regular asymptotic predictions in
Eqs.~\eqref{eq:steadyXm} and \eqref{eq:steadyYp}.
They describe the final rotating state; the transient curves
are obtained from BEM and the time-dependent unexpanded model.
At the scale of Fig.~\ref{fig:bemomega}, the nine histories have
very similar growth, overshoot, and relaxation. Their steady speeds
carry the smaller orientation-dependent differences:
the BEM interaction-induced changes are at most about $1.5\%$
of the matched isolated BEM speed. Table~\ref{tab:omega} resolves
these steady differences.

\begin{table}[!htbp]
\caption{Steady mean angular speeds at $E_0/E_c^{(0)}=2$ for
the material in Eq.~\eqref{eq:bemfarfieldmaterial}.
The BEM and unexpanded values are averages after the
transients have decayed; the final column gives the
first-order regular asymptotic prediction. The last row gives the
isolated references: matched BEM calculations give $1.73118$, while
the unexpanded model and regular asymptotics give $\sqrt3$.
Angular speeds are scaled by $\tau_{\mathrm{MW}}^{-1}$.}
\label{tab:omega}
\centering\small\setlength{\tabcolsep}{4pt}
\begin{tabular}{ccrrr}
\hline\hline
$R$ & $\Theta$ & BEM & Unexpanded & \shortstack{Regular\\asymptotics} \\
\hline
4 & $0^\circ$ & 1.72409 & 1.72614 & 1.72611 \\
4 & $45^\circ$ & 1.73526 & 1.73745 & 1.73733 \\
4 & $90^\circ$ & 1.75638 & 1.75774 & 1.75757 \\
6 & $0^\circ$ & 1.72936 & 1.73029 & 1.73029 \\
6 & $45^\circ$ & 1.73265 & 1.73363 & 1.73362 \\
6 & $90^\circ$ & 1.73868 & 1.73963 & 1.73961 \\
8 & $0^\circ$ & 1.73044 & 1.73131 & 1.73131 \\
8 & $45^\circ$ & 1.73182 & 1.73271 & 1.73271 \\
8 & $90^\circ$ & 1.73435 & 1.73524 & 1.73524 \\
\hline
\multicolumn{2}{c}{Isolated} & 1.73118 & 1.73205 & 1.73205 \\
\hline\hline
\end{tabular}
\end{table}

\subsection{Effects of pair orientation and separation}
The changes in steady rotation depend on the pair orientation.
To measure the effect of the other sphere, we compare each
pair with its corresponding isolated-sphere state. For BEM,
\begin{equation}
\Delta\overline\Omega_{\mathrm{BEM}}
=\overline\Omega_{\mathrm{pair,BEM}}
-\Omega_{\mathrm{iso,BEM}}.
\label{eq:incrementBEM}
\end{equation}
The isolated calculation uses the same numerical settings as
the pair calculation and gives $\Omega_{\mathrm{iso,BEM}}\simeq1.73118$
for both tested axis directions. The unexpanded model and regular
asymptotics have the analytical isolated reference $\sqrt3\simeq1.73205$.
Each pair is compared with its own reference. For example, at $R=8$
and $\Theta=\pi/4$, the BEM pair speed $1.73182$ exceeds the matched
isolated BEM value, even though it lies below $\sqrt3$.

For this material and field strength, the interaction reduces
the mean speed of the counter-rotating pair aligned with the field
($\Theta=0$), but increases it for the oblique counter-rotating
pair ($\Theta=\pi/4$) and the perpendicular co-rotating pair
($\Theta=\pi/2$). The largest increase occurs for perpendicular
co-rotation. These trends are reproduced by the unexpanded model and the regular
asymptotics.

The signs follow from the electric and hydrodynamic
contributions in Eqs.~\eqref{eq:steadyXm} and
\eqref{eq:steadyYp}. For counter-rotation at $\Theta=0$, the
electric interaction reduces the speed more than the
hydrodynamic interaction increases it. At $\Theta=\pi/4$,
the electric reduction is weaker and the hydrodynamic
increase dominates. For perpendicular co-rotation, both
contributions increase the speed. Thus changing the pair
orientation changes the balance between the two interactions.

At oblique pair angles, the two counter-rotating spheres can have different
angular speeds. For example,
at $R=4$ and $\Theta=\pi/4$, the BEM speeds are $1.72625$
and $1.74426$, giving a mean of $1.73526$. The transverse
dipoles perturb the field component along $\ez$ with opposite
signs on the two particles, so their local fields and rotation
speeds differ. The unexpanded model also predicts this
asymmetry.

As the separation increases, the interaction-induced changes
become smaller, as expected from the $R^{-3}$ dipole and
rotlet disturbances. The first-order regular asymptotic predictions are
already close to the unexpanded-model values over this range
(Table~\ref{tab:omega}). The additional mutual couplings in
the unexpanded model produce only small changes to these
steady speeds.

\section{Conclusions}
We have studied how electric and hydrodynamic interactions change the
onset and rotation of two Quincke spheres at a prescribed pair
configuration. At fixed field above onset, the first interaction
correction to the steady angular speed is $\order(R^{-3})$, while the
rotation axes evolve on the slow interaction time. Near onset, the
weakening of magnitude relaxation makes the transverse dipole correction
comparable to the isolated transverse dipole when
$\lambda_0=\order(R^{-3})$. In this regime the transverse dipole and
angular speed are both $\order(R^{-3/2})$, and their evolution is
governed by the coupled amplitude equations.

The near-onset amplitude equations govern general transverse dipole
perturbations. Their linearization identifies four collective modes,
whose critical fields and associated single-mode rotating branches
are determined together with their stability.
Electric interaction gives a common leading shift of the modal
thresholds, while hydrodynamic coupling splits them. Counter-rotation
of type $X_-$ appears first for $0<\Theta<\Theta_{\mathrm m}$ and
co-rotation of type $Y_+$ for $\Theta_{\mathrm m}<\Theta\leq\pi/2$,
where $\Theta_{\mathrm m}=\cos^{-1}(1/\sqrt3)\simeq54.7^\circ$.
Away from onset, the stability boundary also depends on field strength:
$\cos^2\Theta_*=[3(1+\lambda_0)]^{-1}$.
It is $73.2^\circ$ at twice the isolated critical field and approaches
$\Theta_{\mathrm m}$ in the overlap limit
$\delta\ll\lambda_0\ll1$. This agreement connects the two
asymptotic regimes.

Boundary element simulations provide onset intervals and follow the
evolution of rotation from a small charge perturbation. The simulations
and the unexpanded model show rapid growth, overshoot, and damped
relaxation. For the tested material at twice the isolated critical
field, the interaction slows a counter-rotating pair aligned with the
field and speeds up the oblique counter-rotating and perpendicular
co-rotating pairs. The two spheres in an oblique counter-rotating pair can also rotate at different angular speeds. These effects weaken as the separation grows.
The unexpanded dipole--rotlet model provides a simple way to follow the rotation of both spheres by integrating six coupled equations for their dipole moments. It retains selected higher-order contributions and recovers the leading $\order(R^{-3})$ shifts in all four critical fields. For the nine pair configurations in Table~\ref{tab:omega}, its predictions of the steady mean angular speed agree with the BEM simulations to within \(0.13\%\).

Allowing the particle centres to evolve with the computed translational
velocities would couple rotation to changes in pair separation and
orientation. The analysis can also be extended to larger collections of spheres, connecting
with bulk suspension dynamics~\cite{das2023absence}, and to arrays of
spatially constrained rotors motivated by recent micromotor
experiments~\cite{Braun2026}. Other rigid shapes, including cylinders,
would require their corresponding polarization and hydrodynamic
responses. Incorporating wall-mediated interactions would permit a formal
asymptotic treatment of pairs of Quincke rollers~\cite{Bricard2013,Bricard2015}.
Further extensions include isolated deformable drops~\cite{DasSaintillan2021},
interacting drops~\cite{McDougall2026}, and surfactant-covered
drops~\cite{McDougallSurfactant2026}. For deformable drops, the
solvability analysis must couple shape evolution to the interfacial
flow, which depends on the viscosity ratio. For surfactant-covered
drops, gradients in surfactant concentration also enter the tangential
stress balance through Marangoni stresses.

\section*{Data availability}
Supplementary Material is available. The data supporting this study are available from the
corresponding author upon reasonable request. 

\vspace{0.5cm}
\begin{acknowledgments}
This work was supported by the Leverhulme Trust through Research Project Grant
No.~RPG-2026-185.
\end{acknowledgments}

%

\onecolumngrid
\clearpage
\setcounter{section}{0}
\setcounter{equation}{0}
\setcounter{table}{0}
\setcounter{figure}{0}
\setcounter{secnumdepth}{3}
\renewcommand{\thesection}{S\Roman{section}}
\renewcommand{\thesubsection}{\Alph{subsection}}
\renewcommand{\thesubsubsection}{\thesubsection.\arabic{subsubsection}}
\renewcommand{\theequation}{S\arabic{equation}}
\renewcommand{\thetable}{S\arabic{table}}
\renewcommand{\thefigure}{S\arabic{figure}}
\makeatletter
\renewcommand{\p@subsection}{}
\renewcommand{\p@subsubsection}{}
\makeatother
\renewcommand{\theHsection}{supp.\arabic{section}}
\renewcommand{\theHsubsection}{\theHsection.\arabic{subsection}}
\renewcommand{\theHsubsubsection}{\theHsubsection.\arabic{subsubsection}}
\renewcommand{\theHequation}{supp.\arabic{equation}}
\renewcommand{\theHtable}{supp.\arabic{table}}
\renewcommand{\theHfigure}{supp.\arabic{figure}}
\hypertarget{supplement-start}{}
\pdfbookmark[0]{Supplemental Material}{supplement-title}
\begin{center}
{\Large\bfseries Supplemental Material for\\
Electrohydrodynamic interactions of two Quincke rotors: asymptotic theory and boundary element simulations\par}
\medskip
Debasish Das\\[2pt]\small Department of Mathematics and Statistics, University of Strathclyde,\\
\small Glasgow, United Kingdom
\end{center}
\medskip

\section{Derivation of the dipole--rotlet equations}
\label{sec:modelSM}
This section derives the unexpanded dipole--rotlet equations introduced
in Sec.~\ref{sec:setup}, Eqs.~\eqref{eq:modelE}--\eqref{eq:modelP},
of the main text. The surface-charge balance determines dipole evolution;
the electric and hydrodynamic disturbances then couple the two spheres.
\subsection{Surface-charge conservation}
\label{sec:chargeSM}
This subsection establishes the charge-conservation law underlying
Eq.~\eqref{eq:modelP} and the material charge balance in
Eq.~\eqref{eq:bemcharge} of the main text. The corresponding node
motion is given by Eq.~\eqref{eq:bemnodes}.
We first use dimensional variables. A sphere of radius $a$ has centre
$\boldsymbol x_c(t)$, translational velocity $\boldsymbol U$, and angular
velocity $\Om$. Write $\boldsymbol s=\boldsymbol x-\boldsymbol x_c$ for the displacement
from the centre, with $r=|\boldsymbol s|$ and $\boldsymbol n=\boldsymbol s/a$
on the interface. The material velocity is
\begin{equation}
\boldsymbol V=\boldsymbol U+\Om\times\boldsymbol s.
\end{equation}
Let $q_s$ denote free charge per unit area and let the outward normal current
jump be
\begin{equation}
\llbracket\boldsymbol n\cdot\boldsymbol J\rrbracket=\boldsymbol n\cdot(\boldsymbol J^+-\boldsymbol J^-),
\qquad \boldsymbol J^\pm=\sigma_{1,2}\boldsymbol E^\pm.
\end{equation}
The superscripts $+$ and $-$ denote values at the interface on the fluid
and solid sides, respectively. The jump is exterior minus interior, with
the same outward unit normal in both terms.
With no surface conduction, charge conservation is
\begin{equation}
\frac{D_s q_s}{Dt}+q_s\,\boldsymbol\nabla_s\cdot\boldsymbol V+\llbracket\boldsymbol n\cdot\boldsymbol J\rrbracket=0,
\qquad \boldsymbol\nabla_s=(\I-\boldsymbol n\boldsymbol n)\cdot\boldsymbol\nabla.
\label{eq:conservation}
\end{equation}
Rigid motion preserves interfacial area, so
$\boldsymbol\nabla_s\cdot\boldsymbol V=0$.

In coordinates translating with the centre but not rotating with the
particle, define
$\overline q_s(\boldsymbol s,t)=q_s(\boldsymbol x_c(t)+\boldsymbol s,t)$.
The chain rule gives
\begin{equation}
\left.\partial_t\overline q_s\right|_{\boldsymbol s}
 =\left.\partial_t q_s\right|_{\boldsymbol x}+\boldsymbol U\cdot\boldsymbol\nabla q_s.
\end{equation}
Translation is therefore absorbed into the time derivative, and
Eq.~\eqref{eq:conservation} becomes
\begin{equation}
\partial_t\overline q_s+
 (a\Om\times\boldsymbol n)\cdot\boldsymbol\nabla_s\overline q_s
 +\llbracket\boldsymbol n\cdot\boldsymbol J\rrbracket=0.
\label{eq:centredcharge}
\end{equation}
The rotational surface velocity is divergence free, so the advection term
can equivalently be written as a surface divergence. For the frozen pair geometry studied in the main text, the
centres are not advanced and the retained surface motion is rotational.

\subsection{Evolution of the electric dipole moment}
\label{sec:dipolarSM}
We derive the dipole evolution law in Eq.~\eqref{eq:modelP} of the
main text by retaining the degree-one spherical harmonics of the charge
and potential.
\subsubsection{Potentials and boundary conditions}
\label{sec:potentialsSM}
The boundary conditions used here are the same electric continuity and
charge-jump conditions used in Sec.~\ref{sec:bem} of the main text;
the present calculation applies their dipolar projection.
The bulk phases are charge free, giving $\nabla^2\Phi^\pm=0$, with
$\boldsymbol E^\pm=-\boldsymbol\nabla\Phi^\pm$. The potential is continuous
at $r=a$. Gauss' law and the normal current jump give
\begin{align}
q_s&=\boldsymbol n\cdot(\epsilon_1\boldsymbol E^+
 -\epsilon_2\boldsymbol E^-),\label{eq:gauss}\\
\llbracket\boldsymbol n\cdot\boldsymbol J\rrbracket&=\boldsymbol n\cdot(\sigma_1\boldsymbol E^+
 -\sigma_2\boldsymbol E^-).\label{eq:ohm}
\end{align}
Let $\boldsymbol E_i(t)$ be the dimensional incident field at the centre
of sphere $i$, excluding its own field. It is the sum of the imposed
field $E_0\ez$ and the disturbance from the other sphere. The notation
$\boldsymbol E_i$ is retained after nondimensionalization below. Its
degree-one spherical-harmonic potential is
$-r\boldsymbol n\cdot\boldsymbol E_i$. The corresponding exterior and
interior potentials are
\begin{align}
\Phi_1^+&=-r\boldsymbol n\cdot\boldsymbol E_i
 +r^{-2}\boldsymbol n\cdot\Pp_i,\label{eq:phiplus}\\
\Phi_1^-&=-r\boldsymbol n\cdot\boldsymbol E_i
 +\frac r{a^3}\boldsymbol n\cdot\Pp_i.\label{eq:phiminus}
\end{align}
Here $\Pp_i$ is the dimensional total dipole coefficient in the
potential convention of the main text. Scaling it by $a^3E_0$, as in
Sec.~\ref{sec:setup}, gives the dimensionless $\Pp_i$ used there. Equations~\eqref{eq:phiplus} and
\eqref{eq:phiminus} obey potential continuity identically. In a general
incident field they describe its degree-one projection; higher spatial
harmonics are discarded in the unexpanded dipole--rotlet model.

\subsubsection{Charge density and normal current}
\label{sec:currentSM}
This step expresses the two terms in the charge balance using the
material contrasts defined in Eq.~\eqref{eq:contrasts_tauMW}.
The normal derivatives at the interface are
\begin{align}
\partial_r\Phi_1^+|_a&=-\boldsymbol n\cdot\boldsymbol E_i
 -\frac2{a^3}\boldsymbol n\cdot\Pp_i,\\
\partial_r\Phi_1^-|_a&=-\boldsymbol n\cdot\boldsymbol E_i
 +\frac1{a^3}\boldsymbol n\cdot\Pp_i.
\end{align}
Substitution into Eqs.~\eqref{eq:gauss} and \eqref{eq:ohm} gives
\begin{align}
q_{s,1}&=\frac{\epsilon_2+2\epsilon_1}{a^3}\,
 \boldsymbol n\cdot(\Pp_i-a^3\epsc\boldsymbol E_i),\label{eq:q_s1}\\
\bigl(\llbracket\boldsymbol n\cdot\boldsymbol J\rrbracket\bigr)_1&=\frac{\sigma_2+2\sigma_1}{a^3}\,
 \boldsymbol n\cdot(\Pp_i-a^3\sigc\boldsymbol E_i),\label{eq:J1}
\end{align}
where
\begin{equation}
\epsc=\frac{\epsilon_2-\epsilon_1}{\epsilon_2+2\epsilon_1},\qquad
\sigc=\frac{\sigma_2-\sigma_1}{\sigma_2+2\sigma_1},\qquad
\tau_{\mathrm{MW}}=\frac{\epsilon_2+2\epsilon_1}{\sigma_2+2\sigma_1}.
\end{equation}
The permittivity-controlled polarization enters the free-charge density,
whereas the conductivity-controlled polarization enters the Ohmic current.
This distinction determines the two different field terms in the dipole
evolution equation.

\subsubsection{Rigid rotation of the dipolar charge}
\label{sec:rotationSM}
The following identity supplies the rotational convection term in
Eq.~\eqref{eq:modelP} of the main text.
For a vector $\boldsymbol B(t)$ independent of position on the sphere,
\begin{equation}
\boldsymbol\nabla_s(\boldsymbol n\cdot\boldsymbol B)
 =\frac1a(\I-\boldsymbol n\boldsymbol n)\cdot\boldsymbol B.
\end{equation}
Consequently,
\begin{equation}
(a\Om_i\times\boldsymbol n)\cdot
 \boldsymbol\nabla_s(\boldsymbol n\cdot\boldsymbol B)
 =-\boldsymbol n\cdot(\Om_i\times\boldsymbol B).
\label{eq:rotation}
\end{equation}
Rigid rotation preserves the spherical-harmonic degree, so no higher
harmonic is generated by advecting $q_{s,1}$ in this velocity field. Taking
$\boldsymbol B=\Pp_i-a^3\epsc\boldsymbol E_i$ in
Eq.~\eqref{eq:rotation} evaluates the rotational term in
Eq.~\eqref{eq:centredcharge}.

\subsubsection{Dipole moment equation in a time-dependent incident field}
\label{sec:dimdipoleSM}
We now assemble the dimensional dipole equation and recover
Eq.~\eqref{eq:modelP}, including the local-field derivative.
At fixed $\boldsymbol s$ in the translating frame, the normal is independent of time,
and Eq.~\eqref{eq:q_s1} gives
\begin{equation}
\partial_t q_{s,1}=\frac{\epsilon_2+2\epsilon_1}{a^3}\,
 \boldsymbol n\cdot(\dot{\Pp}_i-a^3\epsc\dot{\boldsymbol E}_i).
\end{equation}
Combining this result with Eqs.~\eqref{eq:J1}, \eqref{eq:rotation}, and
\eqref{eq:centredcharge} gives a scalar identity valid for every
$\boldsymbol n$. Projection onto the degree-one harmonics, or use of
$\int_{S^2}\boldsymbol n\boldsymbol n\,\dd\Omega=(4\pi/3)\I$, then gives
\begin{equation}
\dot{\Pp}_i=\Om_i\times(\Pp_i-a^3\epsc\boldsymbol E_i)
 -\frac1{\tau_{\mathrm{MW}}}(\Pp_i-a^3\sigc\boldsymbol E_i)
 +a^3\epsc\dot{\boldsymbol E}_i.
\label{eq:dimdipole}
\end{equation}
The last term arises from differentiating the incident-field contribution
to the free charge. The other sphere's evolving dipole makes this local
field time dependent.

Scaling as in the main text removes $a$ and $\tau_{\mathrm{MW}}$ and yields
\begin{equation}
\dot{\Pp}_i=\Om_i\times(\Pp_i-\epsc\boldsymbol E_i)
 -(\Pp_i-\sigc\boldsymbol E_i)+\epsc\dot{\boldsymbol E}_i.
\label{eq:dimensionlessdipole}
\end{equation}
The isolated critical Mason number is $\Mac$, defined in Eq.~\eqref{eq:Mac}. In dimensional
variables this corresponds to
\begin{equation}
[E_c^{(0)}]^2=\frac{2\mu}
 {\tau_{\mathrm{MW}}\epsilon_1\Mac}.
\end{equation}
The required material condition $\epsc>\sigc$ is equivalent to
$\epsilon_2\sigma_1>\epsilon_1\sigma_2$; for positive conductivities,
it states that the particle charge-relaxation time exceeds that of the fluid.

\subsection{Interaction fields and coupled dipoles}
\label{sec:interactionSM}
This subsection derives the disturbance fields in
Eqs.~\eqref{eq:modelE}--\eqref{eq:modelP} of the main text.
\subsubsection{Electric dipole interaction}
\label{sec:electricSM}
The electric disturbance gives Eq.~\eqref{eq:modelE}.
In the dimensionless variables of the main text, a dipole $\Pp_j$ at
$\boldsymbol x_{c,j}$ produces the potential
\begin{equation}
\Phi_j^d(\boldsymbol x)=\frac{\boldsymbol s_j\cdot\Pp_j}{s_j^3},
\qquad \boldsymbol s_j=\boldsymbol x-\boldsymbol x_{c,j},\quad s_j=|\boldsymbol s_j|.
\end{equation}
Its negative gradient is
\begin{equation}
\boldsymbol E_j^d=-s_j^{-3}
 (\I-3\widehat{\boldsymbol s}_j\widehat{\boldsymbol s}_j)\cdot\Pp_j.
\end{equation}
At the other centre, the local field is therefore
\begin{equation}
\boldsymbol E_i=\ez-\delta\PiR\cdot\Pp_j,\qquad
\PiR=\I-3\widehat{\boldsymbol R}\widehat{\boldsymbol R},\qquad
\delta=R^{-3}.
\label{eq:localfield}
\end{equation}
The dyadic tensor is unchanged on reversing the pair axis, so the same
expression applies to both particles. It has eigenvalues $-2$ along the
pair axis and $1$ in each direction perpendicular to that axis.

\subsubsection{Rotlet interaction and torque balance}
\label{sec:rotletSM}
The rotational response to the fluid disturbance gives Eq.~\eqref{eq:modelO}.
The leading disturbance of a rotating unit sphere is
\begin{equation}
\boldsymbol u_j^d=\frac{\Om_j\times\boldsymbol s_j}{s_j^3}.
\label{eq:rotlet}
\end{equation}
For a constant vector $\Om_j$ and $s_j\ne0$, differentiation gives
\begin{equation}
\boldsymbol\nabla\times\boldsymbol u_j^d
 =-s_j^{-3}(\I-3\widehat{\boldsymbol s}_j\widehat{\boldsymbol s}_j)\cdot\Om_j.
\end{equation}
The electric torque on a sphere in a locally uniform field is
$4\pi\epsilon_1\Pp_i\times\boldsymbol E_i$ in dimensional potential-dipole
units. The viscous torque is
$-8\pi\mu a^3[\Om_i-\tfrac12\boldsymbol\nabla\times\boldsymbol u_e]$,
where $\boldsymbol u_e$ is the dimensional disturbance flow produced
by the other sphere and its vorticity is evaluated at the centre of
sphere $i$. Its dimensionless form is $\boldsymbol u_j^d$ in
Eq.~\eqref{eq:rotlet}, with $j\ne i$.
Using the nondimensional scales, torque balance thus gives
\begin{equation}
\Om_i+\frac\delta2\PiR\cdot\Om_j
 =\Ma^{-1}\Pp_i\times\boldsymbol E_i.
\label{eq:torquebalance}
\end{equation}
The source strength in Eq.~\eqref{eq:rotlet} is the isolated-sphere
angular velocity at leading interaction order. Retaining the coupled
angular velocities in Eq.~\eqref{eq:torquebalance} defines the
unexpanded dipole--rotlet model used here.
\subsubsection{Coupled dipole derivatives}
\label{sec:derivativesSM}
The local-field derivative couples the two dipole rates in Eq.~\eqref{eq:modelP}.
For fixed centres, $\delta$ and $\PiR$ are constant. Differentiating
Eq.~\eqref{eq:localfield} gives
\begin{equation}
\dot{\boldsymbol E}_i=-\delta\PiR\cdot\dot{\Pp}_j.
\end{equation}
The nondimensional dipole equation~\eqref{eq:dimensionlessdipole} can
therefore be written as
\begin{equation}
\dot{\Pp}_i+\epsc\delta\PiR\cdot\dot{\Pp}_j
 =\Om_i\times(\Pp_i-\epsc\boldsymbol E_i)
 -(\Pp_i-\sigc\boldsymbol E_i).
\label{eq:coupleddipole}
\end{equation}
Equations~\eqref{eq:localfield}, \eqref{eq:torquebalance}, and
\eqref{eq:coupleddipole} are the unexpanded dipole--rotlet model used in
Eq.~\eqref{eq:model} of the main text.

To implement the model, let
$\boldsymbol f_i=\Om_i\times(\Pp_i-\epsc\boldsymbol E_i)
 -(\Pp_i-\sigc\boldsymbol E_i)$. The angular velocities and dipole
derivatives follow from two constant-coefficient linear systems:
\begin{align}
\begin{pmatrix}\I&\delta\PiR/2\\\delta\PiR/2&\I\end{pmatrix}
\begin{pmatrix}\Om_1\\\Om_2\end{pmatrix}
 &=\Ma^{-1}\begin{pmatrix}\Pp_1\times\boldsymbol E_1\\
                                  \Pp_2\times\boldsymbol E_2\end{pmatrix},
\label{eq:torquematrix}\\
\begin{pmatrix}\I&\epsc\delta\PiR\\\epsc\delta\PiR&\I\end{pmatrix}
\begin{pmatrix}\dot{\Pp}_1\\\dot{\Pp}_2\end{pmatrix}
 &=\begin{pmatrix}\boldsymbol f_1\\\boldsymbol f_2\end{pmatrix}.
\label{eq:massmatrix}
\end{align}
The torque matrix has eigenvalues $1\pm\delta k/2$, and the derivative
matrix has eigenvalues $1\pm\epsc\delta k$, where $k\in\{-2,1,1\}$.
For positive permittivities and non-overlapping spheres,
$-1/2<\epsc<1$ and $\delta<1/8$, so both matrices are invertible. Computing the two factorizations
once and updating only their right-hand sides gives the
six coupled scalar ordinary differential equations for the three
dipole components of each sphere used in the numerical comparisons.

\section{Phase equilibria and stability away from onset}
\label{sec:outerphaseSM}
This section supplements Sec.~\ref{sec:phaseselection} and
Fig.~\ref{fig:outerselection} of the main text. The counter-rotating
and co-rotating stability conditions are derived there from
Eq.~\eqref{eq:phasecomponents}. Here we examine the additional
branch with $\phi_2=-\phi_1$ and the continuous family at the stability
boundary. The calculation applies away from onset, with
$\delta/\lambda_0\ll1$, after radial and longitudinal transients have relaxed.
We retain $\Pi_{xx}=3\cos^2\Theta-2$ and
$\Pi_{zx}=-3\sin\Theta\cos\Theta$ as in the main text.

To examine the degeneracy, define the half-sum and half-difference
$\phi_{\mathrm s}=(\phi_1+\phi_2)/2$ and
$\phi_{\mathrm a}=(\phi_1-\phi_2)/2$.
Equation~\eqref{eq:phasecomponents} becomes
\begin{align}
\frac{\dd\phi_{\mathrm s}}{\dd\tau}
 &=\frac{\sin\phi_{\mathrm s}}{2}
 [(1-\Pi_{xx})\cos\phi_{\mathrm s}
 -\sqrt{\lambda_0}\Pi_{zx}\cos\phi_{\mathrm a}],\label{eq:SMphis}\\
\frac{\dd\phi_{\mathrm a}}{\dd\tau}
 &=\frac{\sin\phi_{\mathrm a}}{2}
 [(1+\Pi_{xx})\cos\phi_{\mathrm a}
 +\sqrt{\lambda_0}\Pi_{zx}\cos\phi_{\mathrm s}].
\label{eq:SMalphabeta}
\end{align}
The mirror branch $\phi_2=-\phi_1$ has $\phi_{\mathrm s}=0$ and
$\cos\phi_{\mathrm a}=-\sqrt{\lambda_0}\Pi_{zx}/(1+\Pi_{xx})$
when this is admissible. Its two eigenvalues are
$(1-\Pi_{xx}^2+\lambda_0\Pi_{zx}^2)/[2(1+\Pi_{xx})]$ and
$-(1+\Pi_{xx})\sin^2\phi_{\mathrm a}/2$; at least one is positive
whenever the branch exists at a generic oblique angle.
At $\Theta_*$, the bracketed equilibrium conditions in
Eqs.~\eqref{eq:SMphis} and \eqref{eq:SMalphabeta} become dependent:
\begin{equation}
\begin{split}
(1-\Pi_{xx})\cos\phi_{\mathrm s}
 &=\sqrt{\lambda_0}\Pi_{zx}\cos\phi_{\mathrm a},\\
1-\Pi_{xx}^2+\lambda_0\Pi_{zx}^2&=0.
\end{split}
\label{eq:SMphasefamily}
\end{equation}
They give a continuous family connecting the counter-rotating and
co-rotating states. As $\lambda_0\to0^+$, the family tends to
$\phi_{\mathrm s}=\pi/2$, consistent with the continuous family having both $X_-$ and $Y_+$
nonzero in the leading near-onset amplitude equations.

\section{Stability of the near-onset rotating branches}
\label{sec:branchSM}
This section supplements Sec.~\ref{sec:amplitudestability} and
Fig.~\ref{fig:modes}(b) of the main text. The order-by-order
solvability calculation is given in Sec.~\ref{sec:multiple}; here
we give the complete linear stability calculation for the four
single-mode branches. The $X_-$ result reduces to
Eq.~\eqref{eq:branchspectrum} of the main text.
Let $\boldsymbol z_i=(X_i,Y_i)^{\mathsf T}$ and set
$g=\rho-2\sigc(1-3\cos^2\Theta)$. Equations~\eqref{eq:Xamp}
and \eqref{eq:Yamp} of the main text can be written as
\begin{equation}
\frac{\dd\boldsymbol z_i}{\dd T_2}
 =g\boldsymbol z_i+\boldsymbol K\boldsymbol z_j
 -\Mac^{-2}|\boldsymbol z_i|^2\boldsymbol z_i,
\qquad
\boldsymbol K=\operatorname{diag}(-1/2,h).
\end{equation}
Here $h=(2-3\cos^2\Theta)/2$ is the coupling coefficient defined in
Eq.~\eqref{eq:gh} of the main text.
At any equilibrium, the Jacobian in the ordering $(X_1,Y_1,X_2,Y_2)$ is
\begin{equation}
\boldsymbol J^*=\begin{pmatrix}\boldsymbol A_1&\boldsymbol K\\
                              \boldsymbol K&\boldsymbol A_2\end{pmatrix},
\qquad
\boldsymbol A_i=(g-\Mac^{-2}|\boldsymbol z_i|^2)\boldsymbol I_2
 -2\Mac^{-2}\boldsymbol z_i\boldsymbol z_i^{\mathsf T}.
\label{eq:cubicjac}
\end{equation}
For a state with only an $X$ or only a $Y$ dipole component,
perturbations in the two directions decouple. We use the same
sum-and-difference coordinates as Eq.~\eqref{eq:paircoords}, now for
infinitesimal perturbations of a rotating branch:
\begin{equation}
\Delta X_\pm=\frac{\Delta X_1\pm\Delta X_2}{\sqrt2},\qquad
\Delta Y_\pm=\frac{\Delta Y_1\pm\Delta Y_2}{\sqrt2}.
\end{equation}
The columns of Table~\ref{tab:spectra} give the growth rates of equal
($+$) or opposite ($-$) perturbations on the two spheres in the $X$
and $Y$ directions, measured on the slow time $T_2$.

\begin{table}[ht]
\centering\small
\caption{Growth rates about each nonzero single-mode rotating branch,
with $\Lambda_k=\rho+m_k>0$ for that branch. The $X_-$ row gives the growth rates in Eq.~\eqref{eq:branchspectrum}
of the main text.}
\label{tab:spectra}
\begin{tabular}{ccccc}
\hline\hline
Branch & $\Delta X_+$ & $\Delta X_-$ & $\Delta Y_+$ & $\Delta Y_-$\\\hline
$X_+$ & $-2\Lambda_{X_+}$ & $\Lambda_{X_-}-3\Lambda_{X_+}$
 & $\Lambda_{Y_+}-\Lambda_{X_+}$ & $\Lambda_{Y_-}-\Lambda_{X_+}$\\
$X_-$ & $\Lambda_{X_+}-3\Lambda_{X_-}$ & $-2\Lambda_{X_-}$
 & $\Lambda_{Y_+}-\Lambda_{X_-}$ & $\Lambda_{Y_-}-\Lambda_{X_-}$\\
$Y_+$ & $\Lambda_{X_+}-\Lambda_{Y_+}$ & $\Lambda_{X_-}-\Lambda_{Y_+}$
 & $-2\Lambda_{Y_+}$ & $\Lambda_{Y_-}-3\Lambda_{Y_+}$\\
$Y_-$ & $\Lambda_{X_+}-\Lambda_{Y_-}$ & $\Lambda_{X_-}-\Lambda_{Y_-}$
 & $\Lambda_{Y_+}-3\Lambda_{Y_-}$ & $-2\Lambda_{Y_-}$\\\hline\hline
\end{tabular}
\end{table}

For a unique largest mode shift, the corresponding nonzero branch has
negative magnitude and direction growth rates. When the line of centres is parallel to the field, the largest shifts
$m_{X_-}$ and $m_{Y_-}$ coincide. A common change of phase then gives
another equivalent state and has zero growth rate.

At $\Theta=\Theta_{\mathrm m}$, where $\cos^2\Theta=1/3$, the
subspace $X_+=Y_-=0$ is invariant. The remaining steady solutions satisfy
\begin{equation}
X_-^2+Y_+^2=2\Mac^2(\rho+\tfrac12),\qquad \rho>-\tfrac12.
\end{equation}
Thus a continuous family with both $X_-$ and $Y_+$ nonzero connects the
two single-mode branches. Higher-order terms are needed to determine
selection within this family in the unexpanded model.

\section{Critical fields and steady angular velocities}
\label{sec:criticalSM}
This section supplies the supporting calculations for the unexpanded
stationary thresholds in Sec.~\ref{sec:switched} and the regular
asymptotic angular speeds in Sec.~\ref{sec:finiteamp} of the main text.
\subsection{Stationary critical fields of the unexpanded model}
\label{sec:thresholdSM}
We derive and justify the stationary onset fields in
Sec.~\ref{sec:stationaryonset}, Eqs.~\eqref{eq:exactX} and
\eqref{eq:exactY}, before evaluating their crossing.
\subsubsection{Linearized equations for equal and opposite perturbations}
\label{sec:linearSM}
The linearized dipole equations below supplement the neutral
angular-velocity equation~\eqref{eq:neutralmatrix} of the main text.
We retain the coupled dipole--rotlet equations without expanding in
$\delta$ in this linear stability analysis of the non-rotating state.
Their first-order expansion is compared with the near-onset asymptotic
shifts in Eqs.~\eqref{eq:switchedmX} and
\eqref{eq:switchedmY} of the main text.
In Eq.~\eqref{eq:switchedmodel} of the main text, the switches
$s_{\mathrm E}$ and $s_{\mathrm H}$ take values zero or one to disable
or enable electric and hydrodynamic interactions, respectively.
The non-rotating solution is
\begin{equation}
\boldsymbol E_{\mathrm{nr}}
 =(\I+s_{\mathrm E}\sigc\delta\PiR)^{-1}\cdot\ez,
\qquad \Pp_{\mathrm{nr}}=\sigc\boldsymbol E_{\mathrm{nr}}.
\end{equation}
Write $\Pp_i=\Pp_{\mathrm{nr}}+\boldsymbol p_i$,
$\boldsymbol E_i=\boldsymbol E_{\mathrm{nr}}+\boldsymbol e_i$, and
$\Om_i=\boldsymbol\omega_i$. Then
$\boldsymbol e_i=-s_{\mathrm E}\delta\PiR\cdot\boldsymbol p_j$ and
\begin{align}
\boldsymbol\omega_i+s_{\mathrm H}\tfrac\delta2\PiR\cdot\boldsymbol\omega_j
 &=\Ma^{-1}(\boldsymbol p_i\times\boldsymbol E_{\mathrm{nr}}
             +\Pp_{\mathrm{nr}}\times\boldsymbol e_i),\\
\dot{\boldsymbol p}_i+s_{\mathrm E}\epsc\delta\PiR\cdot\dot{\boldsymbol p}_j
 &=\Mac\boldsymbol C(\boldsymbol E_{\mathrm{nr}})\cdot\boldsymbol\omega_i
 -\boldsymbol p_i+\sigc\boldsymbol e_i.
\end{align}
Here $\boldsymbol C(\boldsymbol v)\boldsymbol w=\boldsymbol v\times\boldsymbol w$.
For equal or opposite perturbations, $\boldsymbol p_2=q\boldsymbol p_1$ and
$\boldsymbol\omega_2=q\boldsymbol\omega_1$, with $q=\pm1$. In each
sector let $\boldsymbol p_q=\boldsymbol p_1$, $\boldsymbol e_q=\boldsymbol e_1$,
and $\boldsymbol\omega_q=\boldsymbol\omega_1$. Define
\begin{equation}
\boldsymbol B_q=\I+q s_{\mathrm E}\sigc\delta\PiR,\qquad
\boldsymbol M_q=\I+q s_{\mathrm E}\epsc\delta\PiR,\qquad
\boldsymbol H_q=\I+q s_{\mathrm H}\tfrac\delta2\PiR.
\end{equation}
Eliminating angular velocity gives
\begin{align}
\boldsymbol M_q\dot{\boldsymbol p}_q
 &=(\left(E_0/E_c^{(0)}\right)^2\boldsymbol L_q-\I)\boldsymbol B_q\boldsymbol p_q,
 \qquad \frac{\Mac}{\Ma}=\left(\frac{E_0}{E_c^{(0)}}\right)^2,\label{eq:equaloppositelinear}\\
\boldsymbol L_q
 &=-\boldsymbol C(\boldsymbol E_{\mathrm{nr}})\boldsymbol H_q^{-1}
      \boldsymbol C(\boldsymbol E_{\mathrm{nr}}).
\label{eq:equaloppositeL}
\end{align}
The two exact three-dimensional Jacobians are therefore
\begin{equation}
\boldsymbol J_q=\boldsymbol M_q^{-1}(\left(E_0/E_c^{(0)}\right)^2\boldsymbol L_q-\I)\boldsymbol B_q.
\label{eq:equaloppositeJac}
\end{equation}
In the particle ordering $(\boldsymbol p_1,\boldsymbol p_2)$, the full
six-dimensional Jacobian is
\begin{equation}
\boldsymbol J_6=\frac12
\begin{pmatrix}
\boldsymbol J_++\boldsymbol J_-&\boldsymbol J_+-\boldsymbol J_-\\
\boldsymbol J_+-\boldsymbol J_-&\boldsymbol J_++\boldsymbol J_-
\end{pmatrix}.
\label{eq:sixJac}
\end{equation}

\subsubsection{Reality of the growth rates and stationary onset}
\label{sec:realitySM}
This subsection establishes the reality of the growth rates underlying
the stationary-onset calculation in Sec.~\ref{sec:stationaryonset}.
For sufficiently small $\delta$, the symmetric matrices
$\boldsymbol B_q$, $\boldsymbol M_q$, and $\boldsymbol H_q$ are positive
definite. The first two commute because both are polynomials in $\PiR$.
Define $\boldsymbol u_q=\boldsymbol B_q\boldsymbol p_q
=\boldsymbol p_q-\sigc\boldsymbol e_q$, where
$\boldsymbol e_q=-q s_{\mathrm E}\delta\PiR\cdot\boldsymbol p_q$
is the local-field perturbation. Thus $\boldsymbol u_q$ is the dipole
perturbation minus its conduction polarization in the perturbed local
field. Equation~\eqref{eq:equaloppositelinear} becomes
\begin{equation}
\boldsymbol D_q\dot{\boldsymbol u}_q
 =(\left(E_0/E_c^{(0)}\right)^2\boldsymbol L_q-\I)\boldsymbol u_q,
\qquad \boldsymbol D_q=\boldsymbol M_q\boldsymbol B_q^{-1}.
\label{eq:symmetricJac}
\end{equation}
The mass matrix $\boldsymbol D_q$ is symmetric positive definite, and
the matrix on the right is symmetric. The generalized growth-rate
eigenproblem is therefore equivalent to the symmetric matrix
$\boldsymbol D_q^{-1/2}(\left(E_0/E_c^{(0)}\right)^2\boldsymbol L_q-\I)\boldsymbol D_q^{-1/2}$,
so all growth rates are real. Moreover, $\boldsymbol L_q$ is positive semidefinite because
$\boldsymbol C^{\mathsf T}=-\boldsymbol C$ and $\boldsymbol H_q$ is positive
definite. It has rank two and null vector $\boldsymbol E_{\mathrm{nr}}$.
At $E_0/E_c^{(0)}=0$, the growth operator is negative definite in this metric. The number of its positive eigenvalues
can change only when $\left(E_0/E_c^{(0)}\right)^2$ reaches the reciprocal of a positive eigenvalue
of $\boldsymbol L_q$. The first instability is consequently stationary.

Since $\boldsymbol B_q$ is invertible, the neutral problem can be written
\begin{equation}
\boldsymbol u_q=\left(E_0/E_c^{(0)}\right)^2\boldsymbol L_q\boldsymbol u_q,
\qquad \boldsymbol u_q=\boldsymbol B_q\boldsymbol p_q.
\end{equation}
Neither $\boldsymbol M_q$ nor $\boldsymbol B_q$ enters the resulting
eigenvalues. The derivative coupling therefore affects relaxation rates
but not stationary neutral fields. In particular, the normalized thresholds
are independent of $\epsc$ within the unexpanded dipole--rotlet model.

\subsubsection{Calculation of the four stationary critical fields}
\label{sec:fourfieldsSM}
This calculation evaluates the two mode directions for each exchange
symmetry, yielding Eqs.~\eqref{eq:exactX} and
\eqref{eq:exactY} of the main text.
Take $\widehat{\boldsymbol R}$ and $\boldsymbol E_{\mathrm{nr}}$ in the
$x$--$z$ plane and write
\begin{equation}
F=|\boldsymbol E_{\mathrm{nr}}|^2,\qquad
G=(\widehat{\boldsymbol R}\cdot\boldsymbol E_{\mathrm{nr}})^2.
\end{equation}
The $X_q$ eigenvector of $\boldsymbol L_q$ lies in that plane and is
perpendicular to $\boldsymbol E_{\mathrm{nr}}$. Its cross product with
$\boldsymbol E_{\mathrm{nr}}$ is along $\boldsymbol e_y$, an eigenvector of
$\boldsymbol H_q$ with eigenvalue $1+q s_{\mathrm H}\delta/2$. The $Y_q$ eigenvector is
$\boldsymbol e_y$ itself. The two positive eigenvalues are thus
\begin{equation}
\ell_{X_q}=\frac F{1+q s_{\mathrm H}\delta/2},\qquad
\ell_{Y_q}=\frac{(1+q s_{\mathrm H}\delta/2)F-(3q s_{\mathrm H}\delta/2) G}
 {(1+q s_{\mathrm H}\delta/2)(1-q s_{\mathrm H}\delta)}.
\end{equation}
Taking their reciprocals gives Eqs.~\eqref{eq:exactX} and
\eqref{eq:exactY} of the main text, which are exact within the
unexpanded model.
The in-plane dipole eigenvector is obtained by applying
$\boldsymbol B_q^{-1}$ to the corresponding $\boldsymbol u_q$.

Resolving the imposed field parallel and perpendicular to the pair
axis gives the base field explicitly:
\begin{align}
\boldsymbol E_{\mathrm{nr}}
 &=\frac{\ez-\cos\Theta\,\widehat{\boldsymbol R}}{1+s_{\mathrm E}\sigc\delta}
   +\frac{\cos\Theta\,\widehat{\boldsymbol R}}{1-2s_{\mathrm E}\sigc\delta},\label{eq:baseexplicit}\\
F&=\frac{\sin^2\Theta}{(1+s_{\mathrm E}\sigc\delta)^2}
+\frac{\cos^2\Theta}{(1-2s_{\mathrm E}\sigc\delta)^2},\qquad
G=\frac{\cos^2\Theta}{(1-2s_{\mathrm E}\sigc\delta)^2}.
\label{eq:FG}
\end{align}
Expanding these formulas in $\delta$ gives the common electric shift
$-2s_{\mathrm E}\sigc(1-3\cos^2\Theta)$ and the mode-dependent
hydrodynamic shifts in Eqs.~\eqref{eq:switchedmX} and
\eqref{eq:switchedmY} of the main text. For $s_{\mathrm H}=0$, both
positive eigenvalues equal $F$, so electric interaction alone leaves all
four stationary modal thresholds coincident.

\subsubsection{Crossover angle at finite separation}
\label{sec:crossoverSM}
This calculation supplements Sec.~\ref{sec:crossover} and the kink
in Fig.~\ref{fig:switches}. It converts the local-field condition
in Eq.~\eqref{eq:crosslocal} into the pair angle measured from the
imposed field.
For $s_{\mathrm H}=1$, equating the $X_-$ and $Y_+$ critical fields gives
\begin{equation}
\frac GF=\frac{1+\delta/2}{3(1-\delta/2)}.
\end{equation}
Using Eq.~\eqref{eq:FG}, the pair-axis angle at the crossing is
\begin{equation}
\cos^2\Theta_{\mathrm{cross}}=
\frac{(1+\delta/2)(1-2s_{\mathrm E}\sigc\delta)^2}
 {(1+\delta/2)(1-2s_{\mathrm E}\sigc\delta)^2+2(1-\delta)(1+s_{\mathrm E}\sigc\delta)^2}.
\label{eq:fullcrossclosed}
\end{equation}
For small $\delta$,
\begin{equation}
\cos^2\Theta_{\mathrm{cross}}=\frac13+\frac\delta3-\frac43s_{\mathrm E}\sigc\delta+\order(\delta^2).
\end{equation}
The term $\delta/3$ is the rotlet contribution; the term
$-4s_{\mathrm E}\sigc\delta/3$ is the electric-base-field contribution to the shift. For $\sigc=-1/2$, the
two limits, with angles in radians, are
\begin{equation}
\Theta_{\mathrm{cross}}=\Theta_{\mathrm m}
 -\frac{3\delta}{2\sqrt2}+\order(\delta^2),\qquad
\Theta_{\mathrm{cross}}^{\mathrm H}=\Theta_{\mathrm m}
 -\frac{\delta}{2\sqrt2}+\order(\delta^2).
\end{equation}
These finite-separation displacements of the unexpanded crossover
approach zero with $R^{-3}$. The differences between its modal fields
and the leading near-onset asymptotic predictions are $\order(R^{-6})$.

\subsection{Regular asymptotic steady angular velocities away from onset}
\label{sec:steadyfinite}
This subsection derives the angular-speed corrections in
Eqs.~\eqref{eq:steadyXm} and \eqref{eq:steadyYp} of
Sec.~\ref{sec:finiteamp}, and their overlap with the near-onset theory.
This calculation uses the regular asymptotic expansion away from onset,
at fixed $\lambda_0>0$ with $\delta/\lambda_0\ll1$. We use $\Pperp^{(0)}=\Ma\sqrt{\lambda_0}$ and
$\Pp_i^{(0)}=\Pperp^{(0)}\boldsymbol e_{\rho i}+(\epsc-\Ma)\ez$.
The phase solvability condition $G_{i,\phi}=0$ is imposed before solving the steady
radial and longitudinal equations. Define, only for the following
component calculation,
\begin{equation}
\boldsymbol v_i=\PiR\cdot\Pp_j^{(0)},\qquad
\boldsymbol w_i=\PiR\cdot\boldsymbol e_{\phi j},
\end{equation}
and use $\rho,\phi,z$ for rotating-basis components. The explicit
interaction contribution to the angular velocity is
$\Ma^{-1}[-\Pp_i^{(0)}\times\boldsymbol v_i+(\Pperp^{(0)}/2)\boldsymbol w_i]$.
Substituting this expression into the dipole forcing and solving its
magnitude and longitudinal component equations gives
\begin{align}
P_{i,\rho}^{(1)}
 &=(\Ma-\epsc)v_{i,\rho}
 -\frac{\Ma^2}{\Pperp^{(0)}}v_{i,z}
 +\frac{(\Pperp^{(0)})^2-\Ma^2}{4\Pperp^{(0)}}w_{i,\phi},\label{eq:finitePrho}\\
P_{i,z}^{(1)}
 &=\Pperp^{(0)} v_{i,\rho}-(\Ma+\epsc)v_{i,z}
 -\frac\Ma2 w_{i,\phi}.\label{eq:finitePz}
\end{align}
The phase coordinate is chosen so that $P_{i,\phi}^{(1)}=0$.
For the $X_-$ branch, and for the $Y_+$ branch at $\Theta=\pi/2$,
the angular velocities remain along the leading rotation axes.
Expand the positive angular-speed magnitude as
$|\Om_i|=\sqrt{\lambda_0}+\delta\omega_i^{(1)}+\order(\delta^2)$.
Its first-order coefficient is
\begin{equation}
\omega_i^{(1)}
 =-\frac{(\Pperp^{(0)})^2+\Ma^2}{\Pperp^{(0)}\Ma}
 \left(v_{i,z}+\frac14w_{i,\phi}\right).
\label{eq:finiteomega}
\end{equation}
This identity follows by combining the radial dipole contribution
$P_{i,\rho}^{(1)}/\Ma$ with the explicit interaction torque. In
particular, the terms containing $v_{i,\rho}$ cancel.

For $X_-$, take $\boldsymbol e_{\rho1}=\boldsymbol e_x$ and
$\boldsymbol e_{\rho2}=-\boldsymbol e_x$. Then
$w_{i,\phi}=-1$ and
\begin{equation}
\tfrac12(v_{1,z}+v_{2,z})
 =(\epsc-\Ma)(1-3\cos^2\Theta).
\end{equation}
Averaging the two positive speeds gives
$\overline\Omega_{X_-}$ in Eq.~\eqref{eq:steadyXm} of the main text. For $Y_+$ at $\Theta=\pi/2$,
$\boldsymbol e_{\rho i}=\boldsymbol e_y$,
$w_{i,\phi}=-2$, and $v_{i,z}=\epsc-\Ma$, giving its stated $Y_+$
formula, Eq.~\eqref{eq:steadyYp} of the main text.
Oblique $X_-$ states have individual corrections proportional to
$\sin\Theta\cos\Theta$ with opposite signs in the two speed magnitudes;
these cancel in their mean.

The overlap with the near-onset theory is obtained by taking
$\delta\ll\lambda_0\ll1$. Since
$\Ma=\Mac/(1+\lambda_0)$, the two angular-speed formulas reduce to
\begin{equation}
\overline\Omega_k
 =\sqrt{\lambda_0}
 +\frac{\delta m_k}{2\sqrt{\lambda_0}}
 +\order(\delta\sqrt{\lambda_0})
 +\order(\delta^2\lambda_0^{-3/2}),
\qquad k=X_-,Y_+.
\end{equation}
The same expression follows by expanding
$\sqrt{\lambda_0+\delta m_k}$ from the near-onset amplitude
equations. The last remainder grows as $\lambda_0$ decreases and
becomes comparable to the interaction correction when
$\lambda_0=\order(\delta)$, so this overlap expansion cannot be
continued into the near-onset regime.

\end{document}